%% file: DESY-06-042.tex
\documentclass[zpreprint,zbstdefault]{zeus_paper}
\usepackage[english]{babel}
\input zeus_def.tex
\newcommand{\alnot}{\overline{\alpha_0}}
\newcommand{\alsmzone}{{\alpha_s(M_Z)}}

\newcommand{\pt}{\mbox{$p_T$}}

\newcommand{\Als}{\mbox{$\alpha_{s}$}}
\newcommand{\Alsmz}{\mbox{$\alpha_{s}(M_Z)$}}
\newcommand{\albar}{\mbox{$\overline{\alpha_{0}}$}}
\newcommand{\albarmuI}{\mbox{$\overline{\alpha_{0}}(\mu_{I})$}}
\newcommand{\error}{uncertainty}

\newcommand{\ala}{(\alnot, \alpha_S)}

\newcommand{\mx}{\mathrm{max}}

\def\ee{e^+e^-}

\def\be{\begin{equation}}
\def\ee{\end{equation}}
\def\bea{\begin{eqnarray}}
\def\eea{\end{eqnarray}}

\begin{document}
\include{DESY-06-042-tit}
\include{DESY-06-042-aut}
\include{DESY-06-042-txt}

\include{DESY-06-042-ref}
\include{DESY-06-042-tab}
\include{DESY-06-042-fig}

%
%
\end{document}

%% file: zeus_def.tex
\chardef\usc=95
\chardef\til=126
\catcode`\@=11 
\DeclareRobustCommand\xdotspace{\futurelet\@let@token\@xdotspace}
\def\@xdotspace{%
  \ifx\@let@token.\else
  \ifx\@let@token\bgroup.\else
  \ifx\@let@token\egroup.\else
  \ifx\@let@token\/.\else
  \ifx\@let@token\ .\else
  \ifx\@let@token~.\else
  \ifx\@let@token!.\else
  \ifx\@let@token,.\else
  \ifx\@let@token:.\else
  \ifx\@let@token;.\else
  \ifx\@let@token?.\else
  \ifx\@let@token/.\else
  \ifx\@let@token'.\else
  \ifx\@let@token).\else
  \ifx\@let@token-.\else
  \ifx\@let@token\@xobeysp.\else
  \ifx\@let@token\space.\else
  \ifx\@let@token\@sptoken.\else
   .\space
   \fi\fi\fi\fi\fi\fi\fi\fi\fi\fi\fi\fi\fi\fi\fi\fi\fi\fi}
\catcode`\@=12 

\newcommand{\stru}[2]{%
   \relax\ifmmode\hbox{\vrule height#1 depth#2 width0pt}%
   \else\vrule height#1 depth#2 width0pt\fi}

\newcommand{\Ronum}[1]{\uppercase\expandafter{\romannumeral#1}}
\newcommand{\ronum}[1]{\expandafter{\romannumeral#1}}
\DeclareRobustCommand{\LaTeXZ}{%
  \LaTeX\kern-.05em4\kern-.1em
  {\raisebox{-0.2ex}{$\scriptstyle\text{ZEUS}$}}\xspace}

\newcommand{\eq}[1]{(\ref{eq-#1})}

\newcommand{\Eqsto}[2]{Eqs.~(\ref{eq-#1})--(\ref{eq-#2})}

\DeclareMathAlphabet{\mathbf}{OT1}{cmr}{bx}{sl}
\newcommand{\eVdist}{\kern-0.06667em}

\newcommand{\Gev}{{\text{Ge}\eVdist\text{V\/}}}

\newcommand{\mev}{{\,\text{Me}\eVdist\text{V\/}}}
\newcommand{\gev}{{\,\text{Ge}\eVdist\text{V\/}}}

\newcommand{\Tesla}{\,\text{T}}

\newcommand{\slashfrac}[2]{%
  \raisebox{0.5ex}{\ensuremath #1}\kern-0.12em/\kern-0.08em
  \raisebox{-.8ex}{\ensuremath #2}}

\newcommand{\sqr}[3]{%
    {\vcenter{\hrule height.#3ex\hbox{\vrule width.#2ex height#1ex
     \kern#1ex\vrule width.#3ex}\hrule height.#2ex}}}

\catcode`\@=11 
\newcommand{\parenbar}{\mathpalette\p@renb@r}
\def\p@renb@r#1#2{\vbox{%
  \ifx#1\scriptscriptstyle \dimen@.7em\dimen@ii.2em\else
  \ifx#1\scriptstyle \dimen@.8em\dimen@ii.25em\else
  \dimen@1em\dimen@ii.4em\fi\fi \offinterlineskip
  \ialign{\hfill##\hfill\cr
    \vbox{\hrule width\dimen@ii}\cr
    \noalign{\vskip-.3ex}%
    \hbox to\dimen@{$\mathchar300\hfil\mathchar301$}\cr
    \noalign{\vskip-.3ex}%
    $#1#2$\cr}}}
\catcode`\@=12 

\newcommand{\als}{\alpha_s}

\newcommand{\IP}{{\rm I$\kern-0.01667em$P}\xspace}

\mathchardef\qsm=63
\mathchardef\pls=43
\mathchardef\mns=512
\mathchardef\plm=518
\mathchardef\eql=61
\mathchardef\smallleft=300
\mathchardef\smallright=301
\mathchardef\les=316
\mathchardef\gre=318
\mathchardef\leq=532
\mathchardef\grq=533

\catcode`\@=11 
\newcounter{pict@width}
\newcounter{pict@height}
\newlength{\pict@scale}
\newcommand{\psfigadd}[4]{%
\setcounter{pict@width}{1*\ratio{#2+\pict@scale/2}{\pict@scale}}
\setcounter{pict@height}{1*\ratio{#3+\pict@scale/2}{\pict@scale}}
\setlength{\unitlength}{\pict@scale}
\hbox to #2{\hspace{-\fill}\begin{picture}(\thepict@width,\thepict@height)
\put(0,0){\psfig{figure=#1,width=#2,height=#3,clip=}}
\SetScale{0.283466457}
\SetWidth{1.763889}
{#4}
\end{picture}}
}
\newcounter{pict@widthfst}
\newcounter{pict@widthscd}
\newcounter{pict@widthtot}
\newcommand{\psfigaddtwo}[7]{%
\setcounter{pict@widthfst}{1*\ratio{#2+\pict@scale/2}{\pict@scale}}
\setcounter{pict@widthscd}{1*\ratio{#2+#4+\pict@scale/2}{\pict@scale}}
\setcounter{pict@widthtot}{1*\ratio{#2+#4+#6+\pict@scale/2}{\pict@scale}}
\setcounter{pict@height}{1*\ratio{#3+\pict@scale/2}{\pict@scale}}
\setlength{\unitlength}{\pict@scale}
\hbox{\hspace{-\fill}\begin{picture}(\thepict@widthtot,\thepict@height)
\put(0,0){\psfig{figure=#1,width=#2,height=#3,clip=}}
\put(\thepict@widthscd,0){\psfig{figure=#5,width=#6,height=#3,clip=}}
\SetScale{0.283466457}
\SetWidth{1.763889}
{#7}
\end{picture}}
}
\newcommand{\psfigror}[4]{%
\setcounter{pict@width}{1*\ratio{#2+\pict@scale/2}{\pict@scale}}
\setcounter{pict@height}{1*\ratio{#3+\pict@scale/2}{\pict@scale}}
\setlength{\unitlength}{\pict@scale}
\hbox{\begin{picture}(\thepict@width,\thepict@height)
\put(0,\thepict@height){\psfig{figure=#1,width=#3,height=#2,clip=,angle=270}}
\SetScale{0.283466457}
\SetWidth{1.763889}
{#4}
\end{picture}}
}
\newcommand{\psfigrol}[4]{%
\setcounter{pict@width}{1*\ratio{#2+\pict@scale/2}{\pict@scale}}
\setcounter{pict@height}{1*\ratio{#3+\pict@scale/2}{\pict@scale}}
\setlength{\unitlength}{\pict@scale}
\hbox{\begin{picture}(\thepict@width,\thepict@height)
\put(0,0){\psfig{figure=#1,width=#3,height=#2,clip=,angle=90}}
\SetScale{0.283466457}
\SetWidth{1.763889}
{#4}
\end{picture}}
}
\catcode`\@=12 
\newlength\listtextwidth

\catcode`\@=11 
\newlength{\@tabfninsert}
\newlength{\@tabfnwidth}
\newcommand{\tabfootnote}[2]{%
  \setlength{\@tabfninsert}{0.8em}
  \setlength{\@tabfnwidth}{\textwidth}
  \addtolength{\@tabfnwidth}{-\@tabfninsert}
  \addtolength{\@tabfnwidth}{-0.4em}
  \noindent\makebox[\@tabfninsert][r]{\footnotesize$^{#1}$\hfil}\hfill%
  \parbox[t]{\@tabfnwidth}{\footnotesize #2\hfill}}
\catcode`\@=12 

%% file: DESY-06-042-tit.tex
\prepnum{DESY--06--042}

\title{
Event shapes\\
in deep inelastic scattering\\
at HERA
}                                                       
                    
\author{ZEUS Collaboration}
\draftversion{After Reading}
\date{\today}

\abstract{ 

Mean values and differential distributions of event-shape variables have
been studied in neutral current deep inelastic scattering using an
integrated {luminosity} of 82.2~pb$^{-1}$ collected with the ZEUS
detector at HERA.  The kinematic range  is $80 < Q^2 <
20\,480\gev^2$ and $0.0024 < x < 0.6$, where $Q^2$ is the
virtuality of the exchanged boson and $x$ is the Bjorken variable. The
data are
compared with a model based on a combination of
next-to-leading-order QCD calculations with 
next-to-leading-logarithm corrections 
and the Dokshitzer-Webber non-perturbative power corrections.
The power-correction method provides a reasonable description of the
data for all event-shape variables studied.  Nevertheless, the lack of
consistency of the determination of $\alpha_s$ and of the non-perturbative parameter of the
model, $\albar$, 
suggests the importance of higher-order processes that are
not yet included in the model.

}

\makezeustitle

%% file: DESY-06-042-aut.tex
\pagenumbering{Roman}                                                                              
                                                   %
\begin{center}                                                                                     
{                      \Large  The ZEUS Collaboration              }                               
\end{center}                                                                                       
  S.~Chekanov,                                                                                     
  M.~Derrick,                                                                                      
  S.~Magill,                                                                                       
  S.~Miglioranzi$^{   1}$,                                                                         
  B.~Musgrave,                                                                                     
  D.~Nicholass$^{   1}$,                                                                           
  \mbox{J.~Repond},                                                                                
  R.~Yoshida\\                                                                                     
 {\it Argonne National Laboratory, Argonne, Illinois 60439-4815}, USA~$^{n}$                       
\par \filbreak                                                                                     
  M.C.K.~Mattingly \\                                                                              
 {\it Andrews University, Berrien Springs, Michigan 49104-0380}, USA                               
\par \filbreak                                                                                     
  N.~Pavel~$^{\dagger}$, A.G.~Yag\"ues Molina \\                                                   
  {\it Institut f\"ur Physik der Humboldt-Universit\"at zu Berlin,                                 
           Berlin, Germany}
% ...
\newpage                                                                                           
$^{\    1}$ also affiliated with University College London, UK \\                                  
% ...
$^{\dagger}$ deceased \\                                                                           
%
\newpage   
                                                           %
                                                           %
\begin{tabular}[h]{rp{14cm}}                                                                       
$^{a}$ &  supported by the Natural Sciences and Engineering Research Council of Canada (NSERC) \\  
$^{b}$ &  supported by the German Federal Ministry for Education and Research (BMBF), under        
          contract numbers HZ1GUA 2, HZ1GUB 0, HZ1PDA 5, HZ1VFA 5\\                                
$^{c}$ &  supported in part by the MINERVA Gesellschaft f\"ur Forschung GmbH, the Israel Science   
          Foundation (grant no. 293/02-11.2) and the U.S.-Israel Binational Science Foundation \\  
$^{d}$ &  supported by the German-Israeli Foundation and the Israel Science Foundation\\           
$^{e}$ &  supported by the Italian National Institute for Nuclear Physics (INFN) \\                
$^{f}$ &  supported by the Japanese Ministry of Education, Culture, Sports, Science and Technology 
          (MEXT) and its grants for Scientific Research\\                                          
$^{g}$ &  supported by the Korean Ministry of Education and Korea Science and Engineering          
          Foundation\\                                                                             
$^{h}$ &  supported by the Netherlands Foundation for Research on Matter (FOM)\\                   
$^{i}$ &  supported by the Polish State Committee for Scientific Research, grant no.               
          620/E-77/SPB/DESY/P-03/DZ 117/2003-2005 and grant no. 1P03B07427/2004-2006\\             
$^{j}$ &  partially supported by the German Federal Ministry for Education and Research (BMBF)\\   
$^{k}$ &  supported by RF Presidential grant N 1685.2003.2 for the leading scientific schools and  
          by the Russian Ministry of Education and Science through its grant for Scientific        
          Research on High Energy Physics\\                                                        
$^{l}$ &  supported by the Spanish Ministry of Education and Science through funds provided by     
          CICYT\\                                                                                  
$^{m}$ &  supported by the Particle Physics and Astronomy Research Council, UK\\                   
$^{n}$ &  supported by the US Department of Energy\\                                               
$^{o}$ &  supported by the US National Science Foundation\\                                        
$^{p}$ &  supported by the Polish Ministry of Scientific Research and Information Technology,      
          grant no. 112/E-356/SPUB/DESY/P-03/DZ 116/2003-2005 and 1 P03B 065 27\\                  
$^{q}$ &  supported by FNRS and its associated funds (IISN and FRIA) and by an Inter-University    
          Attraction Poles Programme subsidised by the Belgian Federal Science Policy Office\\     
$^{r}$ &  supported by the Malaysian Ministry of Science, Technology and                           
Innovation/Akademi Sains Malaysia grant SAGA 66-02-03-0048\\                                       
\end{tabular}                                                                                      
                                                           %
                                                           %

%% file: DESY-06-042-txt.tex
\pagenumbering{arabic} 
\pagestyle{plain}

\section{Introduction}
\label{sec-int}
The hadronic final states formed in $e^{+}e^{-}$ annihilation and in
deep inelastic scattering (DIS) can be characterised by a number of
variables that describe the shape of the event.
The event shapes presented in this paper are infrared- and
collinear-safe observables and  can be calculated using
perturbative QCD (pQCD).  In some cases,  the prediction in 
next-to-leading order and next-to-leading-logarithm (NLO+NLL) 
approximation is available.

Precision tests of the pQCD predictions using the experimentally measured
event shapes require a good understanding of  
non-perturbative effects,  namely the hadronisation process, which describes the 
transition from partons to the experimentally observed hadrons.
These non-perturbative corrections decrease as a  power of $Q$,
the square root of the virtuality of the exchanged boson,
and are therefore called power corrections. 
They can be parametrised as $\lambda_p/Q^p$, where the 
scale $\lambda_p$ and exponent $p$ depend on the shape variable;
the exponent $p$ can be predicted by perturbation 
theory~\cite{pl:b339:148,pl:b352:451,np:b437:415,pl:b357:646,*np:b454:253}. 
The success of this simple model in fitting the 
data~\cite{epj:c22:1}
has initiated many studies;
previously non-perturbative effects could be  estimated 
only through the use of Monte Carlo (MC) models.

In the formulation of the power-correction model 
by Dokshitzer and
Webber~\cite{pl:b352:451,pl:b404:321,np:b511:396,*err:np:b593:729,epj:c1:3},  
the shape
variables are given by the sum of the perturbative and non-perturbative
parts which depend only on two constants: the strong coupling,
\Als\, and an effective low-energy coupling, \albar, which is 
universal to all event shapes.  This formulation allows 
the extraction of \Als\ and \albar\ from a fit to the data.

Studies of event shapes at HERA have been already
reported by the 
H1 and ZEUS collaborations ~\cite{epj:c14:255,*hep-ex-0512014,epj:c27:531}.
This paper extends with increased statistics the previous ZEUS 
measurement of the mean event-shape variables to the   
analysis of differential distributions and of two new shape variables.
Measurements were performed in the Breit frame~\cite{feynman:1972:photon,*zfp:c2:237}
in the kinematic range 
$0.0024 < x < 0.6 $, $ 80 < Q^2 < 20\,480\gev^2$ and $ 0.04 < y
< 0.9 $.  Here $x$ is the Bjorken variable and $y = Q^2/sx$, where
$s$ is the centre-of-mass energy squared of the $ep$ system.
Inclusion of the differential distributions allows an improved test of 
the validity
of the power correction method.

\section{Event-shape variables}
\label{sec-es}

The event-shape variables studied in this analysis are thrust, $T$, 
jet broadening, $B$, the invariant jet mass, $M^2$, 
the $C$-parameter, the variable $y_2$ (defined below) and the momentum out of the event
plane, $K_{\rm OUT}$.
Thrust measures the longitudinal collimation of a given hadronic
system, while broadening measures the complementary aspect. These two
parameters are specified relative to a chosen axis, denoted by a
unit vector $\overrightarrow{n}$.  Thus:

\begin{equation}
T=\frac{\sum _{i}\left| \overrightarrow{p_i}\cdot\overrightarrow{n}\right| }{\sum
_{i}\left| \overrightarrow{p_i}\right| } ,
\label{eq-thrust}
\end{equation}
 
\begin{equation}
B=\frac{\sum _{i}\left| \overrightarrow{p_i}\times \overrightarrow{n}\right|
}{\sum _{i}\left| \overrightarrow{p_i}\right| } ,
\label{eq-broad}
\end{equation}
where $\overrightarrow{p_i}$ is  the momentum of the final-state particle $i$.

When $\overrightarrow{n}$ is 
the direction of the virtual-photon, thrust and broadening are denoted by
$T_\gamma$ and $B_\gamma$, respectively.  Alternatively, both
quantities may be measured with respect to the thrust axis, defined as
that direction along which the thrust is maximised by a suitable choice of
$\overrightarrow{n}$.  In this
case, the thrust and broadening are denoted by $T_{T}$ and $B_{T}$.

In the Born approximation,
the final state consists of a single quark,  and
$T_{\gamma}$ and $T_{T}$ are unity.
Consequently, the shape variables $(1 - T_{\gamma})$ and $(1 - T_{T})$
are employed so that non-zero values at the parton level are a direct
indicator of higher-order QCD effects.

The normalised jet invariant mass is defined by

\begin{equation}
M^2 =\frac{ \left(\sum_iE_i\right)^2 - \left| \sum_i \overrightarrow{p_i}
\right| ^2 }{\left( 2\sum _{i}E_i\right)^{2}} ,
\label{eq-jetmass}
\end{equation}

where $E_i$ is  the energy of the final-state particle $i$.

The $C$-parameter is given by   

\begin{equation}
C= \frac{3\sum _{ij}|\overrightarrow{p_{i}}||\overrightarrow{p_{j}}|\sin ^{2}\left( \theta _{ij}\right) }{2(\sum _{i}|\overrightarrow{p_{i}}|)^2} ,
\label{eq-cpar}
\end{equation}
\noindent where $\theta_{ij}$ is the angle between two final-state particles, 
$i$ and $j$.

The shape variables in
\Eqsto{thrust}{cpar} are 
summed over the particles in the current hemisphere of the Breit frame.
To ensure infrared
safety, it is necessary to exclude events in which the energy in the
current hemisphere is less than a certain limit, $\mathcal{E}_{\rm lim}$.
The value $\mathcal{E}_{\rm lim}=0.25\cdot Q$ was used ~\cite{jhep:0002:001}.  The analysis is
based on event shapes calculated in the $P$-scheme, i.e.\ with
particles assumed  
to have zero mass after boosting to the Breit frame. 
  
In addition, two variables, $y_2$ and $K_{\rm OUT}$, 
referred to as two-jet variables, 
are considered.
The quantity $y_2$ is defined as the value of 
the jet resolution cut parameter, $y_{\rm cut}$,
in the $k_{T}$ jet algorithm~\cite{np:b406:187}, at which the transition from (2+1)
to (1+1) jets takes place in a given event; here the first number refers to
the current jet(s) and the second to the proton remnant.

The  variable describing the momentum out of the event plane, $K_{\rm OUT}$,  
has been suggested for study~\cite{jhep:111:66} in events with a configuration
at least as complex as (2+1) jets.
The event plane is
defined by the proton momentum \(\overrightarrow{P}\) in the Breit
frame and the unit vector $\overrightarrow{n}$ which enters the
definition of thrust major:

\begin{equation}
T_M=\max\frac{\sum _{i}\left| \overrightarrow{p_i}\cdot\overrightarrow{n}
\right| }{\sum_{i}\left| \overrightarrow{p_i}\right| }, 
\label{eq-thrustmaj}
\end{equation}

with the additional condition
$\overrightarrow{P}\cdot\overrightarrow{n} = 0$.

The variable $K_{\rm OUT}$ is given by

\begin{equation}
{K_{\rm OUT}}={\sum_{i}\left| {p^{\rm out}_{i}}\right|},
\label{eq-kout}
\end{equation}

where \( {p^{\rm out}_{i}}\) is the component of
momentum $\overrightarrow{p_i}$ of the hadron \(i\) perpendicular to 
the event plane. For leading-order (LO)
(2+1) configurations, since both jets lie in the event plane, 
only non-perturbative effects contribute 
to $K_{\rm OUT}$. At higher orders 
of $\als$, 
perturbative effects will  also 
contribute to  $K_{\rm OUT}$.

In contrast to the definitions in
\Eqsto{thrust}{cpar} the sums in Eqs. \eq{thrustmaj} and \eq{kout} run over all particles 
in the Breit frame.

\section{Experimental set-up}
\label{sec-det}
The data used in this analysis were collected during the 1998-2000 running
period, when HERA operated with protons of energy
$E_p=920$~GeV and electrons or positrons\footnote{In the
following, the term ``electron'' denotes generically both the
electron ($e^-$) and the positron ($e^+$).}
of energy $E_e=27.5$~GeV, and correspond to an integrated luminosity
of $82.2\pm 1.9$~pb$^{-1}$. A detailed description of the ZEUS detector
can be found elsewhere~\cite{pl:b293:465,zeus:1993:bluebook}. A brief
outline of the components that are most relevant for this analysis is
given below.

Charged particles are measured in the central tracking detector
(CTD)~\cite{nim:a279:290,*npps:b32:181,*nim:a338:254}, which operates in a
magnetic field of $1.43\Tesla$ provided by a thin superconducting
solenoid. The CTD consists of 72~cylindrical drift chamber
layers, organised in nine superlayers covering the
polar-angle\footnote{The 
ZEUS coordinate system is a right-handed Cartesian system, with the $Z$ axis 
pointing in the proton beam direction, referred to as the 
``forward direction'', and the $X$ axis pointing left towards the centre of 
HERA. The coordinate origin is at the nominal interaction point.} 
region \mbox{$15^\circ<\theta<164^\circ$}. The transverse momentum
resolution for full-length tracks can be parameterised as
$\sigma(p_T)/p_T=0.0058p_T\oplus0.0065\oplus0.0014/p_T$, with $p_T$ in
$\Gev$. The tracking system was used to measure the interaction vertex
with a typical resolution along (transverse to) the beam direction of
0.4~(0.1)~cm and also to cross-check the energy scale of the calorimeter.

The high-resolution uranium-scintillator calorimeter
(CAL)~\cite{nim:a309:77,*nim:a309:101,*nim:a321:356,*nim:a336:23} covers
$99.7\%$ of the total solid angle and consists of three parts: the
forward (FCAL), the barrel (BCAL) and the rear (RCAL) calorimeters. Each
part is subdivided transversely into towers and longitudinally into one
electromagnetic section and either one (in RCAL) or two (in BCAL and
FCAL) hadronic sections. The smallest subdivision of the calorimeter
is called a cell. Under test-beam conditions, the CAL single-particle
relative energy resolutions were $\sigma(E)/E=0.18/\sqrt{E}$ for
electrons and $\sigma(E)/E=0.35/\sqrt{E}$ for hadrons, with $E$ in GeV.

The luminosity was measured from the rate of the bremsstrahlung process
$ep\rightarrow e\gamma p$. The resulting small-angle energetic photons
were measured by the luminosity
monitor~\cite{desy-92-066,*zfp:c63:391,*acpp:b32:2025}, a
lead-scintillator calorimeter placed in the HERA tunnel at $Z=-107$ m.

\section{Kinematics and event selection}
\label{sec-sel}

A three-level trigger system was used to select events online 
\cite{zeus:1993:bluebook,proc:chep:1992:222}.
Neutral current DIS events were selected by requiring that a scattered 
electron candidate with an energy more than 4~$\gev$ was measured in the 
CAL \cite{epj:c11:427}.

The offline kinematic variables $Q^2$, $x$ and $y$ 
were reconstructed 
using the double angle (DA) method \cite{proc:hera:1991:23}. For offline
selection the electron ($e$)
and the Jacquet-Blondel (JB) \cite{proc:epfacility:1979:391} methods were also
used. 

The offline selection of DIS events was based on the following requirements:

\begin{itemize}
\item $E_e^\prime \gre 10\gev$, where $E_e^\prime$ is the scattered electron energy after correction for energy loss in inactive material in front 
of the CAL, to achieve a high-purity sample of DIS events;

\item $y_e \les 0.9$, where $y_e$ is $y$ as reconstructed by the electron method, to reduce the photoproduction background;

\item $y_{\rm JB} \gre 0.04$, where $y_{\rm JB}$ is $y$ reconstructed by the JB
method, to ensure sufficient accuracy for the DA reconstruction of $Q^2$;

\item $38 \les \delta \les 60\gev$, where $\delta = \sum_i(E-P_Z)_i$ and the sum runs over all CAL energy deposits. The lower cut removed background from
photoproduction and events with large initial-state QED radiation. 
The upper cut removed cosmic-ray background. For events with forward electrons
with $\Theta^e_{\rm lab} < 1$~radian, where the $\Theta^e_{\rm lab}$ is the polar angle in
the laboratory system, 
the $\delta$ cut was tightened to $44 < \delta < 60\gev$, 
to reduce the contributions from  
 electromagnetic deposits outside the CTD that are
likely to be neutral pions wrongly identified as electrons; 

\item $|Z_{\rm vtx}| \les 50$~cm, where $Z_{\rm vtx}$ is the $Z$ position of the
reconstructed primary vertex, to select events consistent with $ep$
collisions.

\end{itemize}

The kinematic range of the analysis is:

\begin{center}
$80<Q^2<20\,480\gev^2$, $0.0024 < x < 0.6$ and $0.04<y<0.9$.
\end{center}

For each event, the reconstruction of the shape variables and jets was performed using a combination of track and CAL information,
excluding the cells and the track associated with the scattered electron. 
The selected tracks and CAL
clusters were treated as massless Energy Flow Objects (EFOs)
\cite{briskin:phd:1998}.
The minimum transverse momentum, $p_{T}$, of each EFO was required to be greater than $0.15$~GeV.

The variables $M^2$, $C$, $T$, and $B$
were reconstructed only using objects in the current region of the Breit frame,
with the following additional requirements:
\begin{itemize}
\item number of EFOs (hadrons, in the case of theoretical calculations) in the current region of the Breit frame $\geq
2$;
\item $|\eta^{\rm EFO}_{\rm lab}| < 1.75$, where $\eta_{\rm lab}$ is
the 
pseudorapidity of an EFO as measured in the laboratory frame.
\end{itemize}


Jets
were reconstructed using the $k_T$ cluster algorithm \cite{np:b406:187} in the 
longitudinally invariant inclusive mode \cite{pr:d48:3160}. 
The jet search was conducted in the entire Breit 
frame. 
For the $y_{2}$ variable,
at least two  EFOs (hadrons) had to be found in the Breit 
frame. Since the proton remnants were explicitly treated by the jet algorithm,
all hadrons from the current and target hemispheres of the Breit frame were
considered.

The $K_{\rm OUT}$ variable
was reconstructed in the entire Breit frame, with the following
cuts required by theory~\cite{jhep:111:66}:
$\eta^{\rm EFO (hadrons)}_{\rm Breit} < 3$, to remove the proton remnants,  and
$y_{2} > 0.1$, to avoid small values of $T_{\rm M}$. In addition, 
$|\eta^{\rm EFO}_{\rm lab}| < 2.2$ was required to select a region of 
well understood acceptance.

\section{Monte Carlo simulation}
\label{sec-MC}

A Monte Carlo  event simulation was used to correct
the data for acceptance and resolution effects.  The detector
simulation was performed with the {\sc Geant}~3.13 program~\cite{tech:cern-dd-ee-84-1}.

Neutral current DIS events were generated using the {\sc Djangoh}~1.1
package~\cite{cpc:81:381}, combining the {\sc Lepto} 6.5.1~\cite{cpc:101:108}
generator with the {\sc Heracles}~4.6.1 program~\cite{cpc:69:155}, which
incorporates first-order electroweak corrections.  The parton cascade
was modelled with the colour-dipole model (CDM), using the
{\sc Ariadne}~4.08 \cite{cpc:71:15} program. In this model, coherence
effects are implicitly included in the formalism of the parton
cascade.  The Lund string-fragmentation model \cite{prep:97:31},
as implemented in {\sc Jetset}~7.4~\cite{cpc:46:43,cpc:82:74}, was used
for the hadronisation phase. 

Additional samples were generated with the {\sc Herwig}~5.9 program
\cite{cpc:67:465}, which does not apply electroweak radiative corrections.
The coherence effects in the final-state cascade are included by
angular ordering of successive parton emissions, and a cluster model
is used for the hadronisation \cite{np:b310:461}. 
Events were also generated using the MEPS option of {\sc Lepto} within {\sc Djangoh},
which subsequently uses a parton showering model similar to {\sc Herwig}.



For {\sc Ariadne}, the default parameters were used.  The {\sc Lepto} simulation
was run with soft-colour interactions turned off, and {\sc Herwig} was
tuned\footnote{The parameter PSPLT was set equal to 1.8; otherwise
default parameters were used.} to give closer agreement with the
measured shape variables at low $Q$; the CTEQ4D~\cite{pr:d55:1280}
parameterisations of the proton parton distribution functions (PDFs)
were used.  The MC event samples were passed through
reconstruction and selection procedures identical to those of the
data. The set of MCs used here ensures that the influence of both
the parton level ({\sc Ariadne} versus {\sc Lepto}, {\sc Herwig}) 
and the fragmentation
({\sc Herwig} versus {\sc Ariadne}, {\sc Lepto}) on the 
systematic uncertainties
is   included.

  The generated distributions include the products of strong and
electromagnetic decays, together with $K^0_S$ and $\Lambda$ decays,
but exclude the decay products of weakly decaying particles with
lifetime greater than \mbox{$3\times10^{-10}$~s}.

\section{QCD calculations}
\label{sec-NLONLL}

\subsection{Perturbative QCD calculations}
\label{sec:pQCD}

The mean values and differential event-shape distributions were analysed using 
different perturbative QCD calculations.  

For the mean event shapes, NLO QCD calculations have
been performed using the programs {\sc DISASTER++}~\cite{hep-ph-9710244} and
{\sc DISENT}\cite{np:b485:291}, 
which give parton-level distributions.  
To determine the theoretical
$\alpha_{s}$ dependence of the variables,  both
programs  were run with  the CTEQ4A  proton PDFs with five 
$\alpha_s$ sets~\cite{pr:d55:1280}.  The mean
value of each shape variable was found to be linearly dependent on
$\Als(M_{\rm Z})$ in the range 0.110$-$0.122. 
The calculations were performed with the renormalisation 
and factorisation scales ${\mu_R}=x_{R}Q$ and 
${\mu_F}=x_{F}Q$, respectively, where for the central analysis
 $x_{R}$ and $x_{F}$ were set to 1.

Infrared and collinear safety ensures that the mean values of 
event shapes can be  computed with fixed order 
calculations~\cite{pl:b339:148,pl:b352:451}. 
However, in order
to describe the differential distributions in the phase space
region where the perturbative radiation is suppressed
(region of small values of the shape parameters), large logarithmic
terms must be resummed.


To obtain the theoretical predictions for 
the differential distributions,  {\sc DISASTER++}
events were generated using the {\sc DISPATCH}~\cite{jhep:0208:032} program
with the MRST99 \cite{epj:c14:133} PDFs. 
The final predictions of the differential distributions,
combining the NLO and NLL calculations as well
as the power corrections (see Section~\ref{sec:pc}), were made using
the  {\sc DISRESUM}~\cite{jhep:0208:032} package described below.

To calculate the perturbative part of the 
differential distribution, $({d\sigma}/{dV})_{\rm PT}$,
where $V$ is the event-shape variable, 
{\sc DISRESUM} matches the NLL resummed perturbative calculation of the differential 
distribution to the corresponding NLO
distribution. The details of the resummed calculations depend on
the type of shape variable i.e global, $T_{\gamma}$ and $B_{\gamma}$,
or non-global, $M$, $C$ and $T_T$  
~\cite{epj:c24:213,pl:b512:323,jhep:0002:001,jhep:0312:007,jhep:0208:032}.
Three  possible types of matching were investigated:
 ${\rm logR}$ matching, similar to that
used in  $e^+ e^-$ annihilation analyses,  ${\rm M}$ and  ${\rm M2}$  matchings.  
The last two were
 specifically  introduced for DIS processes~\cite{epj:c24:213}.
In addition, a modified matching technique can be used for the three types
of matching. 
The modification to the matching ensures firstly that the integrated cross section
has the correct upper limit at $ V = V_{\rm max}$, where
$V_{\rm max}$ is the maximum of the distribution~\cite{epj:c24:213},
and secondly that,
if the fixed order distribution goes smoothly to zero
at the upper limit, the matched-resummed distribution has similar behaviour.
The modification requires that the ${\rm ln}(1/V)$ terms in the resummation are replaced by
expressions of the form 

\begin{equation}
    \frac{1}{p} \ln \left[
    \left(\frac{1}{V}\right)^p -
    \left(\frac{1}{V_\mx}\right)^p + 1\right]\,.
\end{equation}

In addition, to ensure the correct upper limit to the distribution after
non-perturbative corrections, as discussed in Section~\ref{sec:pc},
the shift to the distribution is multiplied by 
\begin{equation}
1 - \left(\frac{V}{V_\mx}\right)^{p_s}.
\end{equation}
The resummation can be expressed in terms 
of a rescaled variable, $1/x_LV$, instead of $1/V$, 
where $x_L$ is a logarithmic rescaling factor~\cite{jhep:0312:007}.
The values of $p$, $p_s$ and $x_L$  were set by default to 1, 2 and 1,
respectively, and 
were varied, as explained in Section~\ref{sec:sys}, to estimate the theoretical
uncertainties of the method.

\subsection{Non-perturbative QCD calculation: power corrections}
\label{sec:pc}

Before the data are compared to the pQCD predictions, the latter  
require correction for the effects of hadronisation.
Dokshitzer and Webber calculated power corrections to the 
event-shape variables in $e^{+}e^{-}$ annihilation, assuming
an infrared-regular behaviour of the effective coupling,
$\alpha_\mathrm{eff}$ 
\cite{pl:b352:451,pl:b404:321,np:b511:396,*err:np:b593:729,epj:c1:3}.  
The
technique was subsequently applied to the case of
DIS~\cite{epj:c1:539} and has been used here.

In this approach, a constant, \albar,  is introduced,
which is independent of the choice of the shape variable.  This constant is
defined as the first moment of the effective strong coupling below the
scale $\mu_{I}$ and is given by: 

\begin{equation}
\albarmuI = \frac{1}{\mu_{I}} \int_{0}^{\mu_{I}}
      \alpha_{\mathrm{eff}}(\mu)d\mu, 
\end{equation} 

where $\mu_{I}$ corresponds to the lower limit where the perturbative
approach is valid. This is taken to be $2\gev$, as in the
previous analyses
~\cite{zfp:c73:229,*pl:b456:322,epj:c1:461,*pl:b459:326,epj:c22:1,epj:c14:255,epj:c27:531}.

The theoretical prediction for the mean values of  an event-shape variable,
denoted by $\left< V\right>$, is then given by

\begin{equation} \left< V\right> (\Als, \albar) =
\left< V\right>_\mathrm{PT}(\Als) + \left< V\right>_\mathrm{pow}(\Als,\albar),
\label{eqn:nlo+power}
\end{equation}

where $\left< V\right>_\mathrm{PT}$ is calculated using the NLO QCD
calculation,
and $\left< V\right>_\mathrm{pow}$ is the
power correction.  The power correction is given by

\begin{equation}
\left< V\right>_\mathrm{pow} = a_{V} \frac{4{\cal{M}}A_{1}}{{\pi}Q}.
\label{eqn:power} 
\end{equation} 

The values of $a_V$ for $(1 - T_T),$  $(1 - T_\gamma),$ $C$
and $M^2$ are respectively 2, 2, $3\pi$ and 1.  For $B_\gamma$  
and
$B_{\rm T}$ 
more complex expressions were used ~\cite{epj:c1:3,epj:c24:213}.

The variable $\cal{M}$ is the `Milan
 factor' of value 1.49~\cite{jhep:9805:3}, which takes into account two-loop
 corrections; it has a relative uncertainty of about $\pm20\%$, due to
 three- and higher-loop effects. The term $A_{1}$ is given by:

\begin{equation}
     A_{1} = \frac{C_{F}}{\pi}\mu_{I} \left[ \overline{\alpha_{0}} -
            \alpha_{s}(\mu_R) - \frac{\beta_{0}}{2\pi} \left(
            \log{\left(\frac{\mu_R}{\mu_{I}}\right)} +
            \frac{K}{\beta_{0}} + 1 \right) \alpha_{s}^{2}(\mu_R)
            \right], \label{eqn:A_1} \end{equation}
where $C_{F}=\frac{4}{3}$, $K=\frac{67}{6} - \frac{\pi^{2}}{2} -
\frac{5}{9}N_{f}$, $\beta_{0}=11-\frac{2}{3}N_{f}$ and $N_{f}$, the number of active flavours,
 is taken to be five. 

The power-corrected differential distributions are given by:

\begin{equation}
\frac{d\sigma}{dV} (V) = \frac{d\sigma}{dV}_{\rm PT} (V - \left<
V\right>_\mathrm{pow}),
\label{eq:diff}
\end{equation}

where ${d\sigma}/{dV}_{\rm PT}$ is calculated as described in the previous
section.

\section{Analysis method}
\label{sec-method}

The event shapes were evaluated for event samples in selected bins of
$x$ and $Q^2$.  The choice of the bin 
sizes~\cite{thesis:hanlon:2004,*thesis:everett:2006} was motivated
by the need to have good statistics and keeping  the
migrations, both between bins, and from the current to the target region
within each bin, small.   The kinematic bin boundaries are listed in
Table~\ref{tab1}. 

The predictions that combine the pQCD calculations and the power corrections
are fitted to the measured mean and differential distributions, with
the exceptions of $y_2$ and $K_{\rm OUT}$, to extract the 
$(\albar, \Als)$
values.  The theoretical predictions for $y_2$ and $K_{\rm OUT}$ are not yet
available,  therefore, no attempt is made to extract $(\albar, \Als)$
from these variables, but they are compared to the NLO QCD calculations
and MC predictions at parton and hadron levels.

Separate $\chi^2$-fits to the mean values as a function of $Q$ and to 
the differential distributions in bins of $Q^2$ are performed  
for each variable. 
The distributions,
when calculated to NLO, diverge at small values of the
shape variable. The divergence is removed when evaluating the integral to 
determine the mean values. Consequently 
for the mean values, the fixed-order NLO prediction is used
combined with the power correction according to Eq.~(\ref{eqn:nlo+power}).
For the differential distributions, the divergence is 
removed by matching NLO to NLL using {\sc DISRESUM}. The distributions
are corrected for hadronisation following Eq.~(\ref{eq:diff}). 
For each observable, the fit was performed  
with  $\Als(M_{\rm Z})$ and  
$\albar$ taken as free parameters.

The fits to both the mean values and the differential 
distributions were made using the Hessian method
~\cite{pr:d65:014012,*jp:g28:779} which uses a full error matrix that includes correlated
off-diagonal terms due to the systematic uncertainties. Therefore, 
the statistical
and systematic uncertainties are not  quoted separately and appear
as one `fit error' in all the tables.

The mean values of the event shapes were evaluated over the
full measured kinematic range. 
The range used in the  fits to the 
differential distribution has been
defined individually for each shape variable and each $Q^2$ range.
The ranges are limited
by the requirements that the pQCD predictions should be
well defined within the bin used in the fit and that the range
used should not extend above the LO upper
limit for the variable.
The first requirement was based on the ratio
{(NLL + NLO + power corrections)/(NLL +NLO)};
bins were omitted   at low values of the shape variable,
where the ratio showed a rapid fall, indicating that  
the power correction is
not well defined in this region. Also  the range
$0.8 < V/V_{\rm max} < 1$ was excluded from the fit
for $1-T_\gamma$, $B_\gamma$ and $M$, where the LO upper limit
is equal to $V_{\rm max}$,  
to avoid the region where theoretical predictions are sensitive to
the details of the matching between NLO and NLL calculations
(where the matching  modification discussed in Section~\ref{sec:pQCD}
 has a large effect).

The final ranges are summarised in Table~\ref{tab2}.
Since the theoretical predictions for differential distributions 
 are reliable only at high 
values of $Q^2$, the fit was restricted to $Q^2 > 320~{\rm GeV^2}$.

\section{Corrections}
\label{sec-unc}

In each ($x$, $Q^2$) bin, the {\sc Ariadne} MC was used to correct for the
event acceptance and the acceptance in each bin of each event-shape
variable.  The acceptance is defined as the
ratio of the number of reconstructed and selected events to the number
of generated events in a given bin.  The acceptance generally exceeds
$70\%$ for all bins, except at extremes of the $Q^2$ range and at low $y$.

Agreement was found between the uncorrected data and the predictions
of {\sc Ariadne} throughout the entire kinematic range of each event-shape
variable (see Section 10.1),
thus confirming its suitability for the purpose of
correcting the data.  
The
data were also compared with the {\sc Herwig} predictions; here the
agreement with data was satisfactory but slightly worse
than when using {\sc Ariadne}.
The correction factors 
were evaluated as the ratios of the generated to the observed values 
in each ($x,Q^2$) bin.  
The correction
procedure accounts for event migration between ($x,Q^2$) intervals,
QED radiative effects, EFO-reconstruction efficiency and energy
resolution, acceptances in \pt\ and $\theta$, and EFO migration between
the current and target regions.
These correction factors
are all within 15\% of unity, and the majority lie within 10\% for the mean
values of the shape variables.
The correction factors for the differential distributions
are typically within 20\% of unity.

\section{Systematic uncertainties} \label{sec:sys}

A detailed study of the sources contributing to the systematic 
uncertainties
of the measurements has been
performed.
The main sources contributing to
the systematic uncertainties are listed below:

\begin{itemize}
\item the data were corrected using a different hadronisation and
parton-shower model, namely {\sc Herwig} or {\sc Lepto}, instead of {\sc Ariadne};
\item the cut $y_e$ was changed from 0.9 to 0.8;
\item the cut on $y_{\rm JB}$ was increased from 0.04 to 0.05;
\item the cut on $\delta$ was tightened from  $38(44) < \delta < 60$ GeV to 
$40(46) < \delta < 60$ GeV; the
harder cut was used to estimate any residual uncertainties in the
photoproduction background;
\item the measured energies of clusters in the calorimeter were varied
by $\pm3\%$, $\pm1\%$ and $\pm2\%$ for the FCAL, BCAL and RCAL, 
respectively, corresponding to the uncertainties of the
associated energy scales;
\item the EFO cuts on $\eta_{\rm lab}$ and $\pt > 150
\mev$  were tightened to $|\eta_{\rm lab}| < 1.5$ and $\pt >
200\mev$; the cuts were also removed.
\end{itemize}

The largest systematic \error\  arose from the choice of  {\sc Herwig} as the
hadronisation model. The other significant systematic was due to the
$\eta_{\rm lab}$ selection. 
The remaining systematics were  smaller
than or similar to the statistical uncertainties.

To estimate the theoretical uncertainties for both the
mean values and the differential distributions, the
renormalisation scale was varied by a factor of two, and studies were
made of the effects of changes to $\mu_{I}$ and to the Milan factor.
To give an indication of the uncertainties due to mass effects, the
data were reanalysed using the $E$-scheme.
For the mean values, the CTEQ4 PDFs were replaced by the MRST99 set.
For the differential distributions, the additional parameters $p$ and $p_s$,
that  ensure the correct behaviour of the  matching
and shift, were varied as shown in Tables~\ref{tab5} and ~\ref{tab6}. The logarithmic rescaling factor, $x_L$, was changed to 1.5
\cite{jhep:0312:007} and the CTEQ5 PDF was used instead of MRST99.

\section{Results}
\label{sec-res}
\subsection{Mean values}

The mean values of the event-shape variables are compared with the {\sc Ariadne} predictions
in Fig.~\ref{fig1new}. 
In general, there is a good agreement between data and MC. 
However the MC tends to overestimate the shape variables at low $Q^2$, in
particular $M^2$. 
The {\sc Ariadne} predictions at the parton level are also shown. 
The difference
between the hadron and parton level demonstrates the contribution from the
hadronisation process, as implemented in {\sc Ariadne}. 
It should be noted that
the parton level of {\sc Ariadne}, defined by the parton shower
model, does not have a rigorous meaning in pQCD~\cite{npps:71:66}
 and should be
taken as indicative only. The structure in the theoretical distributions 
results from 
the different $x$-ranges  associated with the $Q^2$ bins,
see Table~\ref{tab1}.

The mean values of the event-shape variables $(1-T_T)$, $B_T$, $M^2$, $C$,
$(1-T_\gamma)$ and $B_\gamma$ as a function of $Q$
were fitted, by varying $\Als$ and $\alnot$, 
to the sum of an NLO term   obtained from {\sc DISASTER++},
plus the power correction as given by Eq.~(\ref{eqn:power}). 
The data and fit results are shown in Fig.~\ref{fig1}.
For all variables
the theory fits  the data well. For $(1-T_\gamma)$,  the best fit
results in a negative power correction, whereas theory predicts a 
power correction equal to that found for  $(1-T_T)$.

The extracted values $\ala$ are shown in
Fig.~\ref{fig2} and in Tables~\ref{tab3} and \ref{tab4}. The contours on the 
plot represent 
one standard deviation errors, corresponding to about $30\%$ confidence level
(CL), as well as the
$95\%$ CL regions based on the fit errors 
as calculated using the Hessian method. The theoretical uncertainties are 
not shown but are given in the tables, since they result in a correlated shift to all fit results.
The current world average, 
$\alpha_s(M_Z)=0.1182\plm0.0027$~\cite{hep-ex-0407021},
is also shown.

The $\alnot$ values  are in good agreement with  
those previously published ~\cite{epj:c27:531},
but somewhat lower than those obtained by the H1 and  $e^+ e^-$
experiments.
The values of $\Als$ obtained from fits 
to $(1-T_T)$, $B_\gamma$, $C$ and $M^2$ are roughly  consistent with 
each other, but somewhat above the world average value.
However, 
Tables~\ref{tab3} and \ref{tab4} show that the 
theoretical uncertainties are substantial
and strongly correlated between variables.
Fits to $B_T$ and $(1-T_\gamma)$
give values of $\Als$ which are inconsistent with the 
values obtained with the other variables, 
as already observed in the earlier ZEUS measurement.

For  $\Als$, the  dominant uncertainty is that  due to variation of the 
renormalisation scale. For $\albar$, the variation in the Milan factor gives the
largest uncertainty except in the case of $B_\gamma$.
The PDF uncertainty was evaluated by replacing the CTEQ4 PDFs  by the MRST99 set.
With the exception of $(1-T_\gamma)$, the changes in the fitted  $\ala$ are of the order of the 
Hessian fit error. For $(1-T_\gamma)$, the power correction becomes positive and the fitted
values of  $\Als$($\alnot$) change to 0.1285(0.3541), values that are in closer agreement with the 
other variables.
If the model were robust, the fitted 
values of  $\Als$ would be independent of  $\mu_{I}$. 
However a dependence on $\mu_{I}$ is clearly evident in the tables.
In view of these results, no 
attempt to extract combined values of $\ala$ from the mean event
shapes was made.

\subsection{Differential distributions}
\label{sec-res-diff}

The differential distributions of the event-shape variables for  
$Q^2 > 320~{\rm GeV}^2$ 
are compared to the predictions of {\sc Ariadne} in Figs.~\ref{fig2new} and \ref{fig2newb}.
For all variables, {\sc Ariadne} describes the data well.
The parton level of {\sc Ariadne} is also shown. 
The difference between the hadron
and parton levels can be taken as
illustrative of the hadronisation correction.

The differential distributions for $(1-T_\gamma)$, $B_\gamma$, $M^2$, $C$ and
$(1-T_T)$, for which the theoretical predictions are available,
 have been fitted with NLL + NLO +
PC calculations as shown in Figs.~\ref{fig3} and \ref{fig3b}. 
The solid (dashed) bars show the bins
that were used (unused) in the fit as described in 
Section~\ref{sec-method}.


None of the three matching techniques discussed in Section~6.1 is strongly
preferred theoretically. Although the modification terms should be used to
ensure the correct behaviour of the cross section, all options included in {\sc
DISRESUM} have been used. The results of fits using six
different matching options are shown in  Fig.~\ref{fig4} and Tables~\ref{tab7} and 
\ref{tab8}.
The  $\chi^2$ of the fits does not depend significantly
on the form of matching used.
The M2mod matching has been chosen for this analysis
in view of the minimal dispersion of $\als$ and $\alnot$
for this type of matching.
Tables~\ref{tab7} and \ref{tab8} show that the M2mod and Mmod
matching techniques give fitted  $\als$ and $\alnot$ values that agree,
in general, within the Hessian fit errors; an exception is
$B_\gamma$, which shows a five standard deviation shift in $\albar$ when the matching is changed from M2mod to Mmod.
The logRmod matching
gives larger systematic changes in the fitted  $\als$($\alnot$), of the
order of two (one) standard deviation. In all cases, the unmodified matching schemes,
which are theoretically disfavoured, result in larger shifts
than the corresponding modified matching.  
It can be concluded that matching-scheme uncertainties
are approximately twice the Hessian fit errors.

The results of the fit to the differential distributions using
the M2mod matching scheme are summarised
in Tables~\ref{tab5}~and~\ref{tab6}. 
The model gives a good description of the
differential distributions for the 
global variables $(1-T_\gamma)$ and $B_\gamma$
over a substantial range of the
shape variables; the $\chi^2 / \mathrm{dof}$ of the fit is close
to unity. For the non-global variables, $(1-T_T)$, $C$, $M^2$, 
the fit is less good, with  the $\chi^2 / \mathrm{dof}$
lying  in the range two to four. The fitted $\als$ values are consistent with the
world average. With the exception of $C$, the  $\alnot$ values
are consistent with those obtained from the mean values.
Figure~\ref{fig6} shows that, for the global variables, the fitted values of $\als$ and $\alnot$
are consistent with being independent of the $Q$ range. The non-global variables show a larger
sensitivity to the  $Q$ range, possibly reflecting the poorer $\chi^2$ of the fits.

Tables~\ref{tab5}~and~\ref{tab6} also give the theoretical uncertainties in the 
fitted  $\ala$ values. For $\als$, the dominant theoretical uncertainties 
result from the renormalisation scale and the logarithmic rescale factor.
The power factors in the modification terms also give rise to significant
uncertainties for all variables except $(1-T_\gamma)$. In contrast
to the results found for the mean values, changes in the Milan factor and  $\mu_{I}$ 
have no significant influence on the fitted $\als$.
In general, all systematic uncertainties, with exception of the check on PDFs,
for $\albar$ are large compared to the fit errors 
and comparable in size to those of the mean fits.

An estimate of the influence of the fit range is given in the two final lines of 
Tables~\ref{tab5}~and~\ref{tab6}. This estimate was obtained  by 
changing the fit range by half a bin at low values of the shape variable, 
where the influence of the NLL terms is greatest.
For $\als$, the effect of the change is a few percent for  $(1-T_\gamma)$ 
and $B_\gamma$.
In contrast, the fit values for the non-global variables are significantly dependent
on the fit range. 
A comparison with the differential distributions measured by H1
shows reasonable agreement with this analysis. However, 
it should be noted
that the two analyses differ in the kinematic range of the
fits as well as in many details of the fits.
The value of $\albar$, given by H1,
agrees with the interval, approximately 0.4$-$0.5, obtained in 
this analysis. 
 
As in the case of the mean values, the fitted values of $\ala$ for
the differential distributions are inconsistent with one another, with
the non-global variables, $M^2$, $C$ and $(1-T_T)$,  yielding a lower
$\als$ than the global variables $(1-T_\gamma)$ and $B_\gamma$,
irrespective of the matching scheme used.
The uncertainties due to the fit range and theoretical parameters
preclude  a meaningful determination of the average values
for  $\als$ and $\alnot$ from the fits to the differential distribution.

\subsection{Measurement of $y_2$ and $K_{\rm OUT}$}
\label{sec-YK}

As discussed in Section 2, the analyses of the variable $y_2$ and $K_{\rm OUT}$
were
made in the full phase space of the Breit frame, including both the
current and target regions.  In contrast to the variables discussed previously,
the correction to $y_2$ is expected to fall 
as $1/Q^2$.
Although the general form of the correction is known,  the theoretical
calculations are not yet available.
Consequently, no fit has been made 
but the data have been compared to {\sc Ariadne} and NLO predictions.
The distribution of $y_2$ and the mean of  $y_2$ as a function of $Q$
are shown in Figs.~\ref{fig7}a and \ref{fig7}b, respectively,  
together with the {\sc Ariadne} predictions at the hadron and parton levels.
The figures show that {\sc Ariadne} describes the $y_2$ 
distribution for $Q^2 > 320~{\rm GeV}^2$ well, but overestimates the means at lower $Q$.

In Fig.~\ref{fig7}c, the  $y_2$ distributions 
are compared with the  NLO distribution from {\sc DISENT} calculated using
$\als(M_{\rm Z}) = 0.116$.
Except at the lowest $y_2$ value for high $Q$, the NLO 
predictions describe
the data well.
In Fig.~\ref{fig7}d, the mean values of $y_2$ are plotted as a function of $Q$
and compared with the NLO predictions.
The agreement with the NLO predictions 
is good over the entire range of $Q$. 

The $K_{\rm OUT}$ variable measures the momentum out of the event plane defined by two 
jets and thus depends on $\als^2$ at lowest order. 
The data are compared to {\sc Ariadne}  
predictions at the parton and hadron level  
in Fig.~\ref{fig11}. 
For the differential distribution, at the parton level,
{\sc Ariadne} 
agrees well with the tail of the $K_{\rm OUT}$ distribution but peaks at a
lower value
than the data.  
The hadron-level prediction, on the other hand, 
describes the data well everywhere, indicating the importance of
hadronisation corrections to this variable.
The mean value of  $K_{\rm OUT}/Q$ agrees well with the 
expectation from {\sc Ariadne}  
at the hadron level. In contrast, the parton-level predictions lie 
below the data, with a difference that decreases with $Q$, again indicating the
importance of hadronisation effects.

\section{Summary}

Measurements have been made of mean values and differential distributions
of the event-shape
variables thrust $T$, broadening $B$, normalised jet mass $M^2$,
$C$-parameter,  $y_2$   and $K_{\rm OUT}$ using the ZEUS detector at HERA.  The variables $T$
and $B$ were determined relative to both the virtual photon axis and the
thrust axis.  The events were analysed in the Breit frame for the
kinematic range $ 0.0024 < x < 0.6 $, $ 80 < Q^2 <
20\,480\gev^2$ and $ 0.04 < y < 0.90 $.  The data are well
described by the {\sc Ariadne}  Monte Carlo model.

The $Q$ dependence of the mean event shapes $T$, $B$,  $M^2$
and $C$,
have been fitted to NLO
calculations from perturbative QCD using the {\sc DISASTER++}
program together with the Dokshitzer-Webber non-perturbative power
corrections, with the strong coupling  \Alsmz\
and the effective non-perturbative coupling $\albar$ as free parameters.

Consistent values of \Als\ are obtained
for the shape variables $(1-T_T)$, $B_\gamma$, $M^2$ and $C$, with \albar\ values
that agree to within $\pm10\%$.
For  $B_T$, the \albar\ value agrees with other variables,
whereas \Als\ does not.  The variable $(1-T_\gamma)$ gives   \Als\ and \albar\
values that are  inconsistent with the other variables and, 
in contrast to the other variables,  that are
sensitive to the parton density used in {\sc DISASTER++}.  
For all variables, 
the renormalisation uncertainties, the dominant theoretical uncertainty,
are three to ten times larger than the experimental uncertainties. Also
the $\mu_{I}$
parameter used in the power corrections gives
large uncertainties.  These may be indications for the need for
higher orders in the power corrections.

The program {\sc DISRESUM} together with NLO calculations from {\sc DISPATCH} has been
used to fit the differential distributions for the event-shape variables
$(1-T_T)$, $M^2$, $C$, $(1-T_{\gamma})$ and $B_{\gamma}$ 
for  $Q^2 > 320~{\rm GeV}^2$.
A reasonable description is obtained for all variables. The modified
matching schemes give fitted values of \Als\ that
are consistent with the world average.  With the exception of $C$, the values of \albar\ are consistent with
those found from the mean values
and lie within the range 0.4$-$0.5.

Comparison between the $\Als$ and $\albar$ from the fits to different 
variables show, however, that the results are not consistent within
the experimental uncertainties.  The renormalisation uncertainties are
still large.
There is a considerable sensitivity to small changes in the kinematic
range of fits, indicating problems in the theoretical description of the
data.  Also the choice of matching scheme 
produces variations of the order of the experimental 
 uncertainties.

The  power corrections for the variables  $y_2$   and $K_{\rm OUT}$
are not yet available. For $y_2$, the data are well
described by NLO calculations.
The variable $K_{\rm OUT}$ is well described by {\sc Ariadne}  predictions 
at the hadron level.
A comparison of  $K_{\rm OUT}$ with parton and hadron level predictions of
 {\sc Ariadne} 
indicates the need for substantial hadronisation corrections.

In summary, the power-correction method provides a reasonable description of the
data for all event-shape variables studied.  Nevertheless, the lack of
consistency of the \Als\ and \albar\
 determinations obtained in deep
inelastic scattering, for the mean values in particular,
suggests the importance of higher-order processes that are
not yet included in the model.

\section*{Acknowledgements}
It is a pleasure to thank the DESY Directorate 
for their strong support and encouragement. The remarkable achievements of the
HERA machine group were essential for the successful completion of this work and
are greatly appreciated. The design, construction and installation
of the ZEUS detector has been made possible by the efforts of many people
who are not listed as authors.  
We are indebted to M.~Dasgupta, 
G.~Salam, M.~Seymour, G.~Marchesini, G.~Zanderighi and A.~Banfi
 for many invaluable discussions.

%% file: DESY-06-042-ref.tex
{
\def\bibname{\Large\bf References}
\def\refname{\Large\bf References}
\pagestyle{plain}
\ifzeusbst
  \bibliographystyle{./BiBTeX/bst/l4z_default}
\fi
\ifzdrftbst
  \bibliographystyle{./BiBTeX/bst/l4z_draft}
\fi
\ifzbstepj
  \bibliographystyle{./BiBTeX/bst/l4z_epj}
\fi
\ifzbstnp
  \bibliographystyle{./BiBTeX/bst/l4z_np}
\fi
\ifzbstpl
  \bibliographystyle{./BiBTeX/bst/l4z_pl}
\fi
{\raggedright
\bibliography{./BiBTeX/user/syn.bib,%
              ./BiBTeX/user/ADAM.bib,%
              ./BiBTeX/bib/l4z_articles.bib,%
              ./BiBTeX/bib/l4z_books.bib,%
              ./BiBTeX/bib/l4z_conferences.bib,%
              ./BiBTeX/bib/l4z_h1.bib,%
              ./BiBTeX/bib/l4z_misc.bib,%
              ./BiBTeX/bib/l4z_old.bib,%
              ./BiBTeX/bib/l4z_preprints.bib,%
              ./BiBTeX/bib/l4z_replaced.bib,%
              ./BiBTeX/bib/l4z_temporary.bib,%
              ./BiBTeX/bib/l4z_zeus.bib}
}
\vfill\eject

%% file: DESY-06-042-tab.tex
%
%

\begin{table}[htbp] 
  \begin{center}     
    \begin{tabular}{||c|c|c||c|c|c||}
      \hline
      Bin&$Q^{2}\;(\gev^2$)&$x$\\      
      \hline 
      1 & ~80 $-$ 160 &0.0024 $-$ 0.010 \\
      2 & 160 $-$ 320 & 0.0024 $-$ 0.010 \\       
      3 & 320 $-$ 640 & ~0.01 $-$ 0.05 \\           
      4 & 640 $-$ 1280 & ~0.01 $-$ 0.05 \\           
      5 & 1280 $-$ 2560 & ~0.025 $-$ 0.150 \\           
      6 & 2560 $-$ 5120 & ~0.05 $-$ 0.25 \\           
      7 & 5120 $-$ 10240 & ~0.06 $-$ 0.40 \\       
      8 & 10240 $-$ 20480 & ~0.10 $-$ 0.60 \\ 
      \hline  
    \end{tabular} 
    \caption{The kinematic boundaries of the bins in $x$ and $Q^2$.}      
    \label{tab1}     
  \end{center}
\end{table}

\begin{table}[htbp]
  \begin{center}
    \begin{tabular}{|c||c|c|c|c|c|}
      \hline
      $Q^{2}\;(\gev^2$)&$1-T_{T}$&$M^2$&$C$&$1-T_{\gamma}$&$B_{\gamma}$\\
      \hline\hline
      ${320 - 640}$
      &$0.1 - 0.3 $
      &$0.05 - 0.2 $
      &$0.3 - 0.7 $
      &$ 0.1 - 0.8$
      &$ 0.15 - 0.4$
      \\
      ${640 - 1280}$
      &$ 0.05 - 0.3 $
      &$ 0.025 - 0.2$
      &$ 0.2 - 0.7$
      &$ 0.1 - 0.8 $
      &$ 0.15 -0.4 $
      \\
      $1280 - 2560$
      &$ 0.05 - 0.3 $
      &$ 0.025 - 0.2 $
      &$ 0.2 - 0.7 $
      &$ 0.1 - 0.8 $
      &$  0.1 - 0.4$
      \\
      ${2560 - 5120 }$
      &$ 0.05 - 0.3 $
      &$ 0.025 - 0.2 $
      &$ 0.1 - 0.7 $
      &$ 0.1 - 0.8 $
      &$  0.1 - 0.4 $
      \\
      $ 5120 - 10240$
      &$ 0.05 - 0.3  $
      &$ 0.025 - 0.2  $
      &$ 0.1 - 0.7  $
      &$ 0.1 - 0.8 $
      &$ 0.05 - 0.4  $
      \\
      $  10240 - 20480$
      &$ 0.05 - 0.3 $
      &$ 0.025 - 0.2 $
      &$ 0.1 - 0.7 $
      &$ 0.1 - 0.8   $
      &$ 0.05 - 0.4 $\\
      \hline  
    \end{tabular} 
    \caption{Ranges used for fits to the differential distributions.}  
    \label{tab2}     
  \end{center}
\end{table}


\begin{table}[p]
  \begin{center}
    \begin{tabular}{|c||c|c|c|c|c|c|}
      \hline
      Variable&$1-T_{T}$&$B_{T}$&$M^2$&$C$&$1-T_{\gamma}$&$B_{\gamma}$\\
      \hline\hline
      $\boldsymbol{\alsmzone}$ 
      &$\mathbf{0.1252} $
      &$\mathbf{0.1149} $
      &$\mathbf{0.1231} $
      &$\mathbf{0.1263} $
      &$\mathbf{0.1456} $
      &$\mathbf{0.1231} $
      \\
      \hline\hline
      {\it Fit~error\/}
      &$\pm0.0010 $
      &$\pm0.0008 $
      &$\pm0.0010 $
      &$\pm0.0006 $
      &$\pm0.0035 $
      &$\pm0.0022 $
      \\
      $\chi^2 / dof$
      &$ 0.4150 $
      &$ 0.4873 $
      &$ 1.4003 $
      &$ 0.4127 $
      &$ 0.9725 $
      &$ 2.6992 $
      \\
      {\it correlation } 
      &$ -0.5337 $
      &$ -0.5719 $
      &$ -0.5275 $
      &$ -0.1133 $
      &$ -0.9257 $
      &$  0.7610 $
      \\
      \hline            
      $ x_{R} = 0.5$
      &$-0.0070  $
      &$-0.0068  $
      &$-0.0077  $
      &$-0.0072  $
      &$-0.0095 $
      &$-0.0062  $\\      
      $ x_{R} = 2.0$
      &$ +0.0085 $
      &$ +0.0065 $
      &$ +0.0091 $
      &$ +0.0088 $
      &$ +0.0104  $
      &$ +0.0067 $\\      
      ${\cal{M}} = 1.19$
      &$ +0.0026 $
      &$ +0.0019 $
      &$ +0.0024 $
      &$ +0.0030 $ 
      &$ +0.0034 $
      &$ +0.0012 $\\      
      $ {\cal{M}} = 1.79$
      &$-0.0023  $
      &$-0.0017  $
      &$-0.0021  $
      &$-0.0026  $  
      &$-0.0029  $
      &$-0.0011  $\\      
      $ \mu_{I}=1~\gev$
      &$+0.0056  $
      &$+0.0039  $
      &$+0.0052  $
      &$+0.0067  $  
      &$+0.0075  $
      &$+0.0024  $\\      
      $ \mu_{I}=4~\gev$
      &$-0.0061  $
      &$-0.0047  $
      &$-0.0057  $
      &$-0.0070  $  
      &$-0.0077  $
      &$-0.0032  $\\            
      {\it E-scheme}
      &$+0.0046  $
      &$+0.0031  $
      &$+0.0036  $
      &$+0.0033  $
      &$+0.0033  $
      &$+0.0013  $\\
      {\it PDF}
      &$-0.0010  $
      &$-0.0005  $
      &$-0.0015  $
      &$-0.0008  $
      &$-0.0172  $
      &$-0.0024  $\\
      \hline       
    \end{tabular}
    \caption{Results for $\alpha_{s}(M_Z)$ from the fit
      to the mean values of the shape variables. The fit error 
      is the total experimental error including both  statistical and
      experimental systematic errors. 
      The correlation coefficients are those between the fitted values
      of $\alpha_{s}(M_Z)$ and $\alnot$ (see Table~\protect\ref{tab4}).  The
      theoretical uncertainties (see text) are also shown.
}
    \label{tab3}
  \end{center}
\end{table}
\begin{table}[p]
  \begin{center}
    \begin{tabular}{|c||c|c|c|c|c|c|}
      \hline
      Variable&$1-T_{T}$&$B_{T}$&$M^2$&$C$&$1-T_{\gamma}$&$B_{\gamma}$\\
      \hline\hline
      $\boldsymbol{\overline{\alpha_{0}}}\;\mathbf{(2~GeV)}$ 
      &$\mathbf{0.4622} $
      &$\mathbf{0.4349} $
      &$\mathbf{0.4184} $
      &$\mathbf{0.4122} $
      &$\mathbf{0.2309} $
      &$\mathbf{0.4352} $
      \\
      \hline\hline            
      {\it Fit~error\/}
      &$\pm0.0047 $
      &$\pm0.0044 $
      &$\pm0.0074 $
      &$\pm0.0030 $
      &$\pm0.0167 $
      &$\pm0.0044 $
      \\     
      \hline       
      $ x_{R} = 0.5$
      &$+0.0105  $
      &$+0.0316  $
      &$+0.0239  $
      &$+0.0094  $
      &$+0.0339  $
      &$+0.1625  $\\      
      $ x_{R} = 2$
      &$-0.0036  $
      &$-0.0089  $
      &$-0.0111  $
      &$-0.0039  $
      &$+0.0063  $
      &$-0.1030  $\\      
      ${\cal{M}} = 1.19$
      &$+0.0360 $
      &$+0.0343 $
      &$+0.0258 $
      &$+0.0232 $
      &$-0.0507  $
      &$+0.0272 $\\      
      $ {\cal{M}} = 1.79$
      &$-0.0280  $
      &$-0.0294  $
      &$-0.0210  $
      &$-0.0198  $  
      &$+0.0252  $
      &$-0.0201  $\\            
      {\it E-scheme}
      &$+0.0157  $
      &$+0.0079  $
      &$+0.0120  $
      &$+0.0130  $  
      &$-0.0043  $
      &$+0.0072  $\\
      {\it PDF}
      &$+0.0139  $
      &$+0.0113  $
      &$+0.0169  $
      &$+0.0129  $
      &$+0.1232  $
      &$+0.0032  $
      \\  
      \hline
    \end{tabular}
    \caption{Results for $\albar$ from the fit
      to the mean values of the shape variables. The fit error 
      is the total experimental error including both  statistical and
      experimental systematic errors. The
      theoretical uncertainties (see text) are also shown.}
    \label{tab4}
  \end{center}
\end{table}                               
%
%
%
%
\begin{table}[p]
 \begin{center}
            \begin{tabular}{|c||c|c|c|c|c|}
            \hline 
            Variable&$1-T_{T}$&$M^2$&$C$&$1-T_{\gamma}$&$B_{\gamma}$\\
            \hline\hline 
$\boldsymbol{\alsmzone}$
&$\mathbf{0.1151} $
&$\mathbf{0.1158} $
&$\mathbf{0.1176} $
&$\mathbf{0.1227} $
&$\mathbf{0.1226} $
\\\hline\hline
 {\it Fit~error\/}
&$\pm0.0016 $
&$\pm0.0013 $
&$\pm0.0016 $
&$\pm0.0012 $
&$\pm0.0013 $
\\ 
 \hline
$Mmod$
&$ -0.0009 $
&$ -0.0001 $
&$ +0.0007 $
&$ -0.0011 $
&$ -0.0020 $
\\$M$
&$ -0.0020 $
&$ -0.0040 $
&$ -0.0033 $
&$ -0.0012 $
&$ -0.0031 $
\\$M2$
&$ -0.0038 $
&$ -0.0048 $
&$ -0.0011 $
&$ -0.0025 $
&$ -0.0036 $
\\$logRmod$
&$ -0.0031 $
&$ -0.0032 $
&$ -0.0029 $
&$ -0.0029 $
&$ +0.0012 $
\\$logR$
&$ -0.0045 $
&$ -0.0056 $
&$ -0.0026 $
&$ -0.0040 $
&$ -0.0054 $
\\ 
 \hline 
 \end{tabular} 
    \caption{Results for $\alpha_{s}(M_Z)$ from the fit
      to the differential distributions of the shape variables using the
      $\rm M2mod$ matching scheme. The fit error 
      is the total experimental error including both  statistical and
      experimental systematic errors. 
      The theoretical uncertainties due to the use of different
      matching schemes (see text) are shown.
}
\label{tab7}
\end{center}
\end{table}
\begin{table}[p]
 \begin{center}
            \begin{tabular}{|c||c|c|c|c|c|}
            \hline 
            Variable&$1-T_{T}$&$M^2$&$C$&$1-T_{\gamma}$&$B_{\gamma}$\\
            \hline\hline 
$\boldsymbol{\overline{\alpha_{0}}}\;\mathbf{(2~GeV)}$
&$\mathbf{0.4173} $
&$\mathbf{0.4650} $
&$\mathbf{0.3358} $
&$\mathbf{0.4820} $
&$\mathbf{0.4268} $
\\\hline\hline
 {\it Fit~error\/}
&$\pm0.0134 $
&$\pm0.0100 $
&$\pm0.0138 $
&$\pm0.0138 $
&$\pm0.0217 $
\\ 
 \hline
$Mmod$
&$ -0.0012 $
&$ +0.0013 $
&$ +0.0042 $
&$ +0.0162 $
&$ +0.1048 $
\\$M$
&$ +0.0041 $
&$ +0.0264 $
&$ +0.0496 $
&$ +0.0129 $
&$ +0.1329 $
\\$M2$
&$ +0.0087 $
&$ +0.0287 $
&$ +0.0657 $
&$ +0.0272 $
&$ +0.1506 $
\\$logRmod$
&$ +0.0018 $
&$ +0.0130 $
&$ +0.0165 $
&$ +0.0159 $
&$ +0.0286 $
\\$logR$
&$ +0.0107 $
&$ +0.0335 $
&$ +0.0650 $
&$ +0.0360 $
&$ +0.1707 $
\\  
 \hline 
 \end{tabular} 
    \caption{Results for $\albar$ from the fit
      to the differential distributions of the shape variables using the
      $\rm M2mod$ matching scheme. The fit error 
      is the total experimental error including both  statistical and
      experimental systematic errors. 
      The theoretical uncertainties due to the use of different
      matching schemes (see text) are shown.
}
\label{tab8}
\end{center}
\end{table}

\begin{table}[p]
 \begin{center}
            \begin{tabular}{|c||c|c|c|c|c|}
            \hline 
            Variable&$1-T_{T}$&$M^2$&$C$&$1-T_{\gamma}$&$B_{\gamma}$\\
            \hline\hline 
$\boldsymbol{\alsmzone}$
&$\mathbf{0.1151} $
&$\mathbf{0.1158} $
&$\mathbf{0.1176} $
&$\mathbf{0.1227} $
&$\mathbf{0.1226} $
\\\hline\hline
 {\it Fit~error\/}
&$\pm0.0016 $
&$\pm0.0013 $
&$\pm0.0016 $
&$\pm0.0012 $
&$\pm0.0013 $
\\$\chi^2 / dof$
&$ 3.33 $
&$ 2.46 $
&$ 3.97 $
&$ 0.74 $
&$ 0.50 $
\\{\it correlation }
&$ -0.72 $
&$ -0.72 $
&$ -0.64 $
&$ -0.69 $
&$ -0.75 $
\\ 
 \hline
$ x_{R} = 0.5$
&$ -0.0023 $
&$ -0.0039 $
&$ -0.0039 $
&$ -0.0040 $
&$ -0.0028 $
\\$ x_{R} = 2$
&$ +0.0054 $
&$ +0.0057 $
&$ +0.0051 $
&$ +0.0060 $
&$ +0.0049 $
\\${\cal{M}} = 1.19$
&$ +0.0000 $
&$ -0.0000 $
&$ +0.0003 $
&$ +0.0001 $
&$ +0.0000 $
\\${\cal{M}} = 1.79$
&$ -0.0000 $
&$ +0.0000 $
&$ -0.0002 $
&$ -0.0001 $
&$ -0.0001 $
\\$ \mu_{I}=1~GeV$
&$ +0.0001 $
&$ -0.0001 $
&$ +0.0004 $
&$ +0.0002 $
&$ +0.0001 $
\\$ \mu_{I}=4~GeV$
&$ -0.0001 $
&$ +0.0000 $
&$ -0.0007 $
&$ -0.0004 $
&$ -0.0002 $
\\$ x_{L} = 1.5$
&$ +0.0045 $
&$ +0.0046 $
&$ +0.0049 $
&$ +0.0024 $
&$ +0.0038 $
\\$ x_{L} = 0.67$
&$ -0.0044 $
&$ -0.0048 $
&$ -0.0044 $
&$ -0.0034 $
&$ -0.0062 $
\\$ p = 2.0$
&$ -0.0029 $
&$ -0.0044 $
&$ -0.0069 $
&$ -0.0011 $
&$ -0.0038 $
\\$ p_s = 1.0$
&$ +0.0029 $
&$ +0.0031 $
&$ +0.0025 $
&$ +0.0016 $
&$ +0.0013 $
\\{\it E-scheme}
&$ -0.0049 $
&$ +0.0033 $
&$ -0.0114 $
&$ +0.0009 $
&$ -0.0004 $
\\{\it PDF}
&$ +0.0000$
&$ +0.0003$
&$ +0.0004$
&$ +0.0009$
&$ +0.0008$ 
\\{\it -0.5 Bins}
&$ -0.0143 $
&$ -0.0112 $
&$ -0.0066 $
&$ -0.0019 $
&$ -0.0022 $
\\{\it +0.5 Bins}
&$ +0.0103 $
&$ +0.0073 $
&$ +0.0086 $
&$ +0.0039 $
&$ +0.0037 $
\\ 
 \hline 
 \end{tabular} 
    \caption{Results for $\alpha_{s}(M_Z)$ from the fit
      to the differential distributions of the shape variables. The fit error 
      is the total experimental error including both  statistical and
      experimental systematic errors. 
      The correlation coefficients are those between the fitted values
      of $\alpha_{s}(M_Z)$ and $\alnot$ (see Table~\protect\ref{tab6}).  The
      theoretical uncertainties (see text) are also shown.
}
\label{tab5}
\end{center}
\end{table}
\begin{table}[p]
 \begin{center}
            \begin{tabular}{|c||c|c|c|c|c|}
            \hline 
            Variable&$1-T_{T}$&$M^2$&$C$&$1-T_{\gamma}$&$B_{\gamma}$\\
            \hline\hline 
$\boldsymbol{\overline{\alpha_{0}}}\;\mathbf{(2~GeV)}$
&$\mathbf{0.4173} $
&$\mathbf{0.4650} $
&$\mathbf{0.3358} $
&$\mathbf{0.4820} $
&$\mathbf{0.4268} $
\\\hline\hline
 {\it Fit~error\/}
&$\pm0.0134 $
&$\pm0.0100 $
&$\pm0.0138 $
&$\pm0.0138 $
&$\pm0.0217 $
\\ 
 \hline
$ x_{R} = 0.5$
&$ -0.0419 $
&$ -0.0344 $
&$ -0.0366 $
&$ -0.0444 $
&$ -0.0637 $
\\$ x_{R} = 2$
&$ +0.0114 $
&$ +0.0215 $
&$ +0.0233 $
&$ +0.0284 $
&$ +0.0503 $
\\${\cal{M}} = 1.19$
&$ +0.0453 $
&$ +0.0571 $
&$ +0.0211 $
&$ +0.0505 $
&$ +0.0394 $
\\${\cal{M}} = 1.79$
&$ -0.0301 $
&$ -0.0377 $
&$ -0.0150 $
&$ -0.0349 $
&$ -0.0258 $
\\$ x_{L} = 1.5$
&$ +0.0231 $
&$ +0.0308 $
&$ +0.0289 $
&$ +0.0442 $
&$ +0.0665 $
\\$ x_{L} = 0.67$
&$ -0.0203 $
&$ -0.0230 $
&$ -0.0262 $
&$ -0.0385 $
&$ -0.0299 $
\\$ p = 2.0$
&$ +0.0042 $
&$ +0.0223 $
&$ +0.0424 $
&$ +0.0109 $
&$ +0.1010 $
\\$ p_s = 1.0$
&$ +0.0009 $
&$ +0.0029 $
&$ -0.0098 $
&$ -0.0014 $
&$ +0.0145 $
\\{\it E-scheme}
&$ +0.0006 $
&$ +0.0026 $
&$ +0.0410 $
&$ +0.0107 $
&$ -0.0030 $
\\ {\it PDF}
&$ +0.0009$
&$ -0.0019$
&$ -0.0023$
&$ -0.0035$
&$ -0.0027$
\\{\it -0.5 Bins}
&$ +0.1473 $
&$ +0.1079 $
&$ +0.0751 $
&$ +0.0437 $
&$ +0.0514 $
\\{\it +0.5 Bins}
&$ -0.0021 $
&$ -0.0203 $
&$ -0.1039 $
&$ +0.0348 $
&$ -0.0688 $
\\
 \hline 
 \end{tabular} 
    \caption{Results for $\albar$ from the fit
      to the differential distributions of the shape variables. The fit error 
      is the total experimental error including both  statistical and
      experimental systematic errors. The
      theoretical uncertainties (see text) are also shown.}
\label{tab6}
\end{center}
\end{table}


%% file: DESY-06-042-fig.tex

\begin{figure}[htbp]
\vfill
\begin{center}
{\epsfig{figure=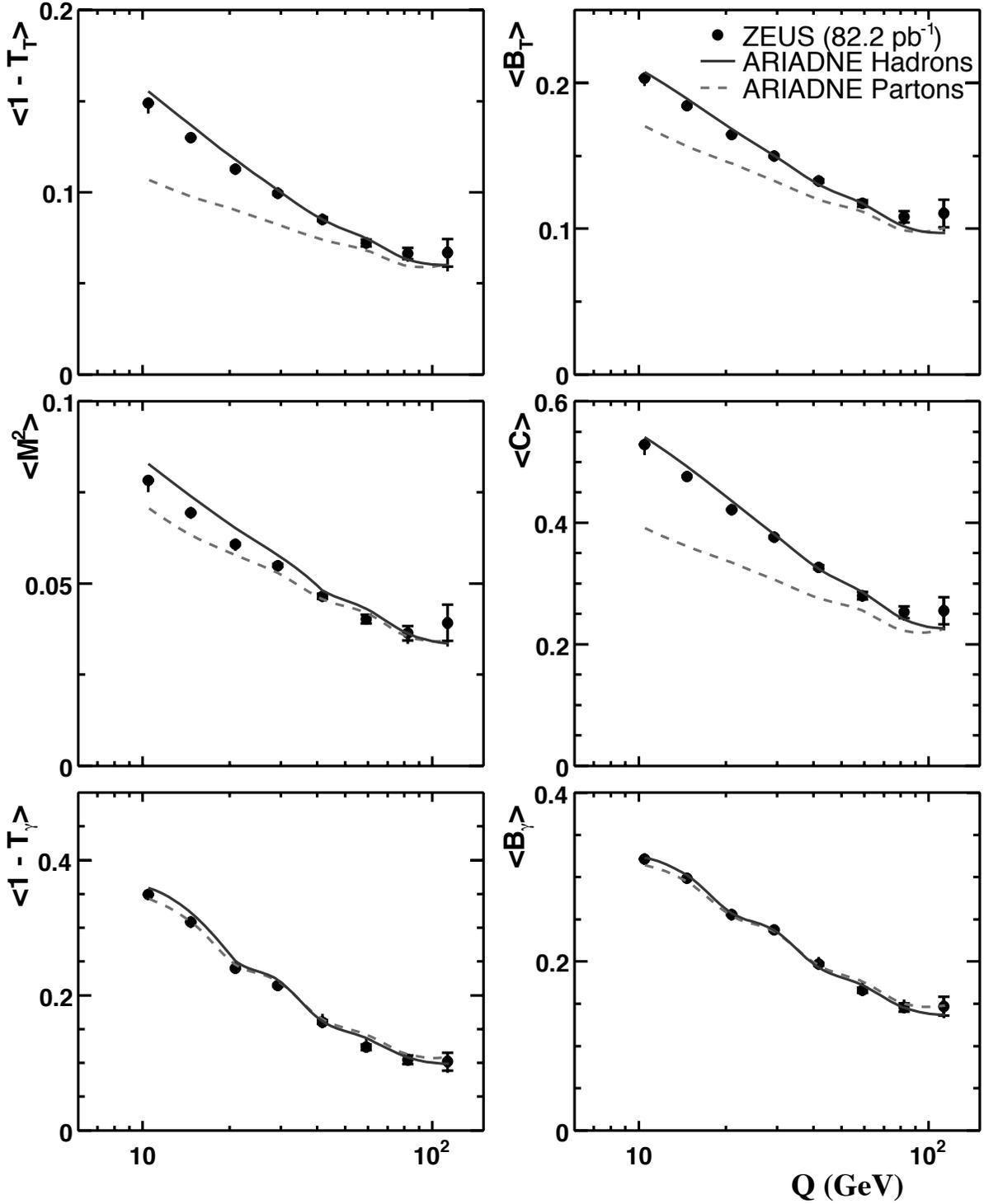,width=\linewidth}}
\end{center}
\caption{
  The mean values of event-shape variables as a function of Q.  
  The solid and dashed lines are the hadron and parton level predictions from 
  ARIADNE, respectively.
}
\label{fig1new}
\vfill
\end{figure}

\begin{figure}[htbp]
\vfill
\begin{center}
{\epsfig{figure=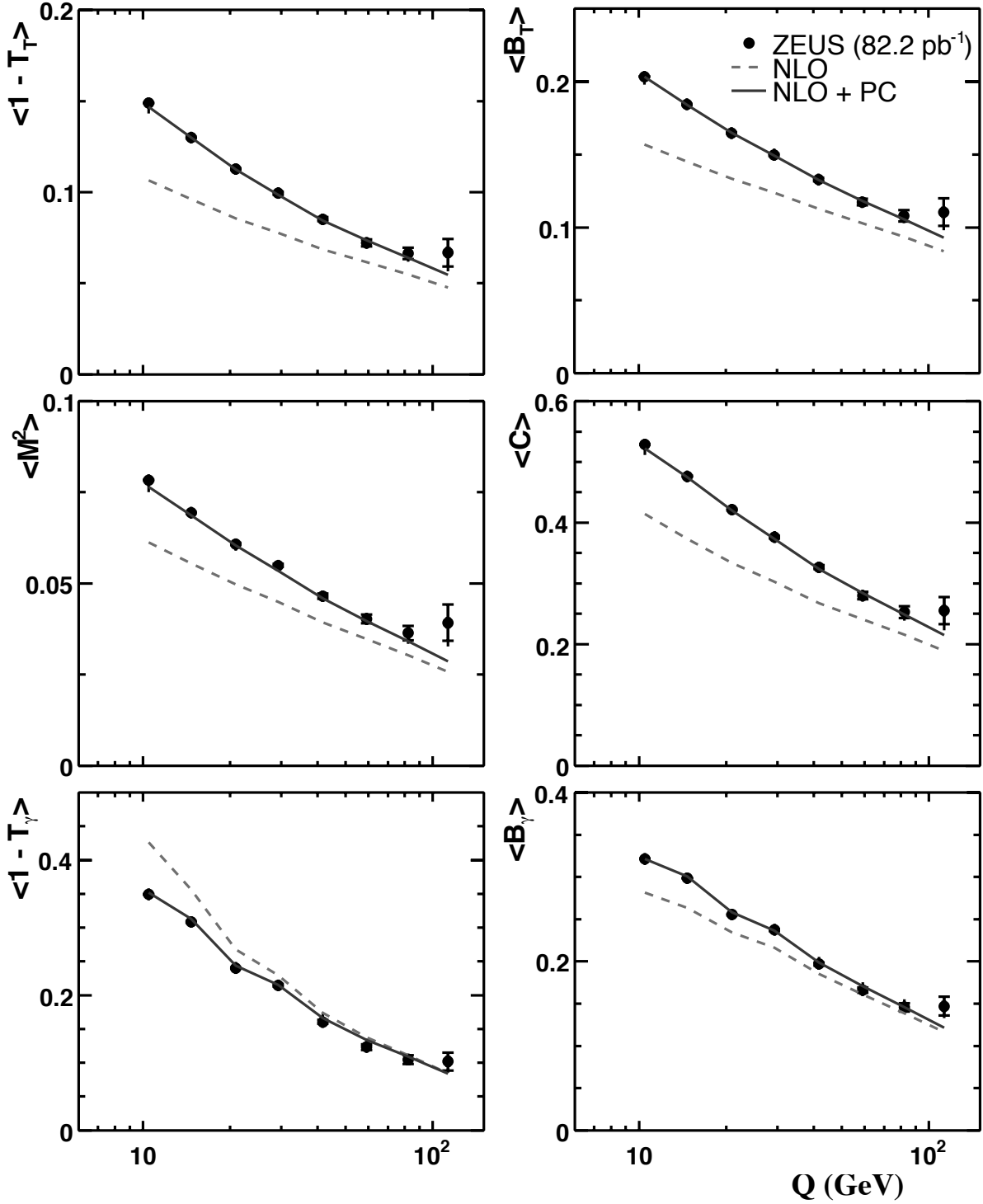,width=\linewidth}}
\end{center}
\caption{
  The mean values of event-shape variables as a function of Q. The solid
  lines are the results of the fit to the data of 
  the predictions of the sum of NLO pQCD
  calculations from {\sc DISASTER++} and the power corrections.  
  The dashed lines are the {\sc DISASTER++} contribution to the fit alone.
}
\label{fig1}
\vfill
\end{figure}

\begin{figure}[htbp]
\vfill
\begin{center}
\centerline{\epsfig{figure=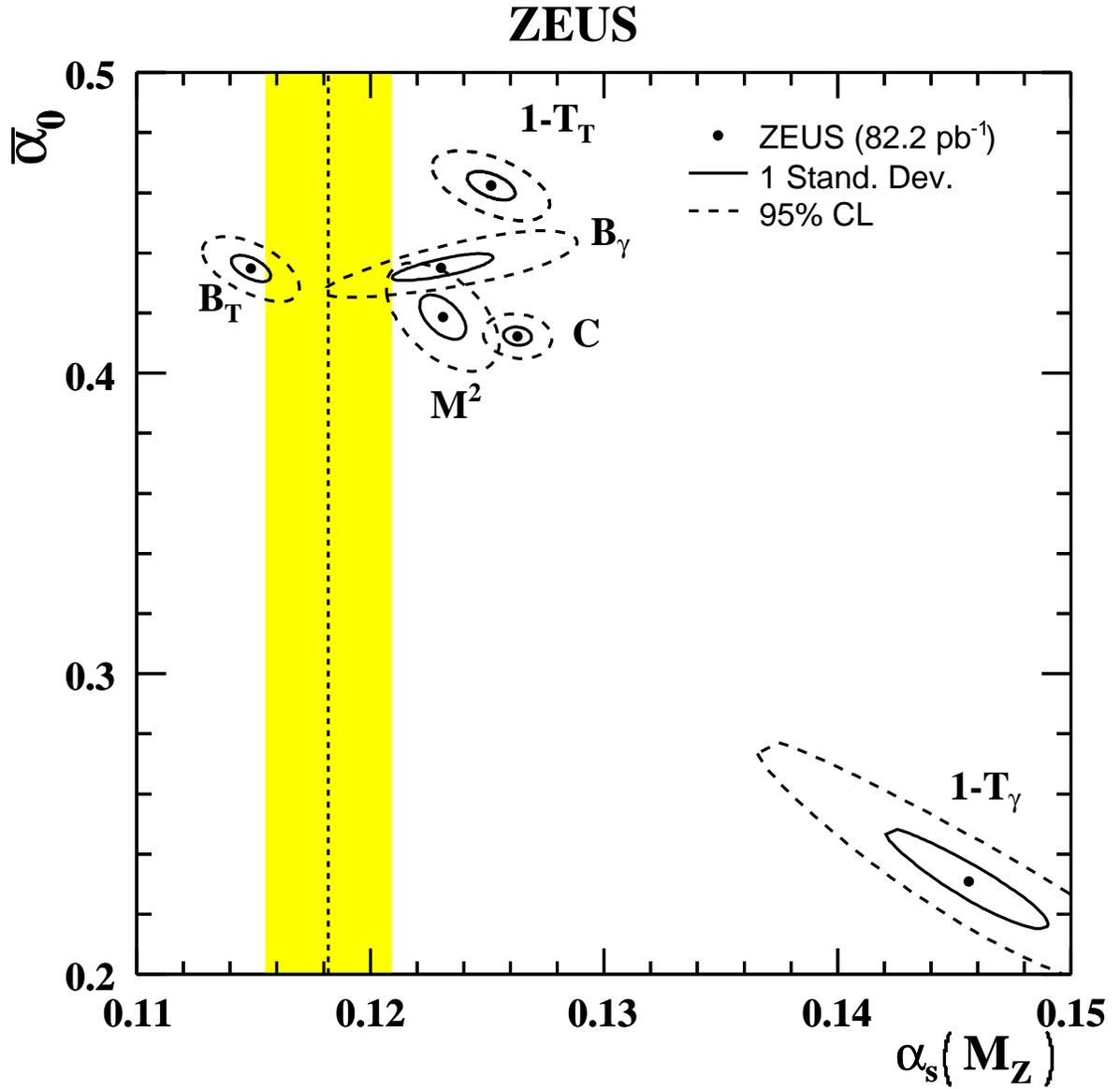,width=\linewidth}}
\end{center}
\caption{
  Extracted parameter values for $(\als,\alnot)$ from fits 
  to the mean values of the shape variables.
  The vertical line and shaded area indicate
  the world average value of $\alsmzone$\protect \cite{hep-ex-0407021}.  }
\label{fig2}
\vfill
\end{figure}

\begin{figure}[htbp]
\vfill
\begin{center}
{\epsfig{figure=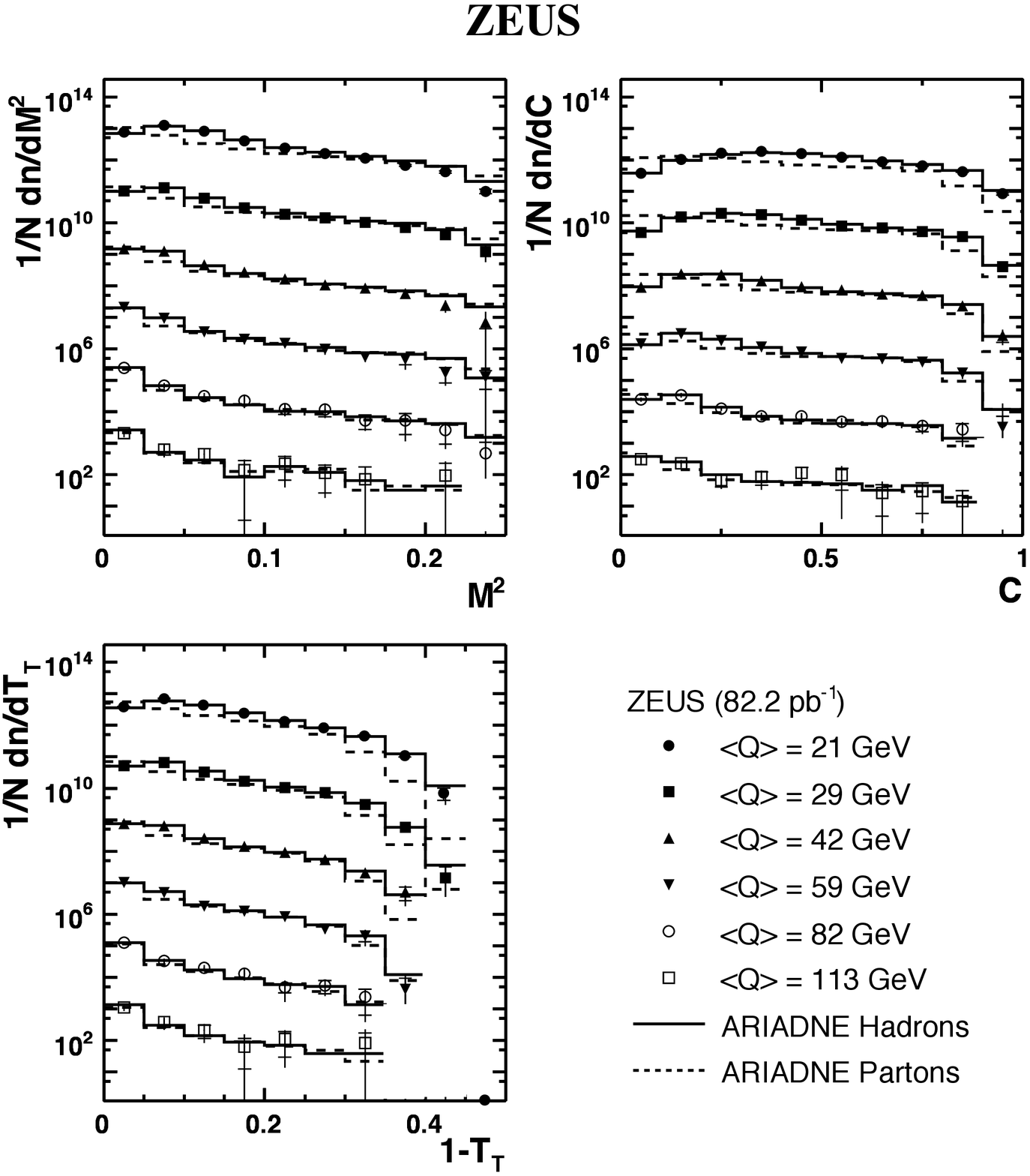,width=\linewidth}}
\end{center}
\caption{Differential distributions for the event shapes 
$M^2$, $C$ and $1-T_T$.
The distributions are normalised such that n refers to
the number of events in the $(x,Q^2)$ bin after the
$\mathcal{E}_{\rm lim}$ cut and N to the total number of
events in the $(x,Q^2)$ bin before the
$\mathcal{E}_{\rm lim}$ cut.
The differential cross section has been scaled 
for clarity by factors $10^n$, where $n$= 12, 10, 8, 6, 4, 2 for 
$<Q>$= 21, 29, 42, 59, 82 and 113 GeV, respectively. 
Predictions of ARIADNE at the hadron (solid lines) and parton
(dashed lines) levels are shown.
}
\label{fig2new}
\vfill
\end{figure}

\begin{figure}[htbp]
\vfill
\begin{center}
{\epsfig{figure=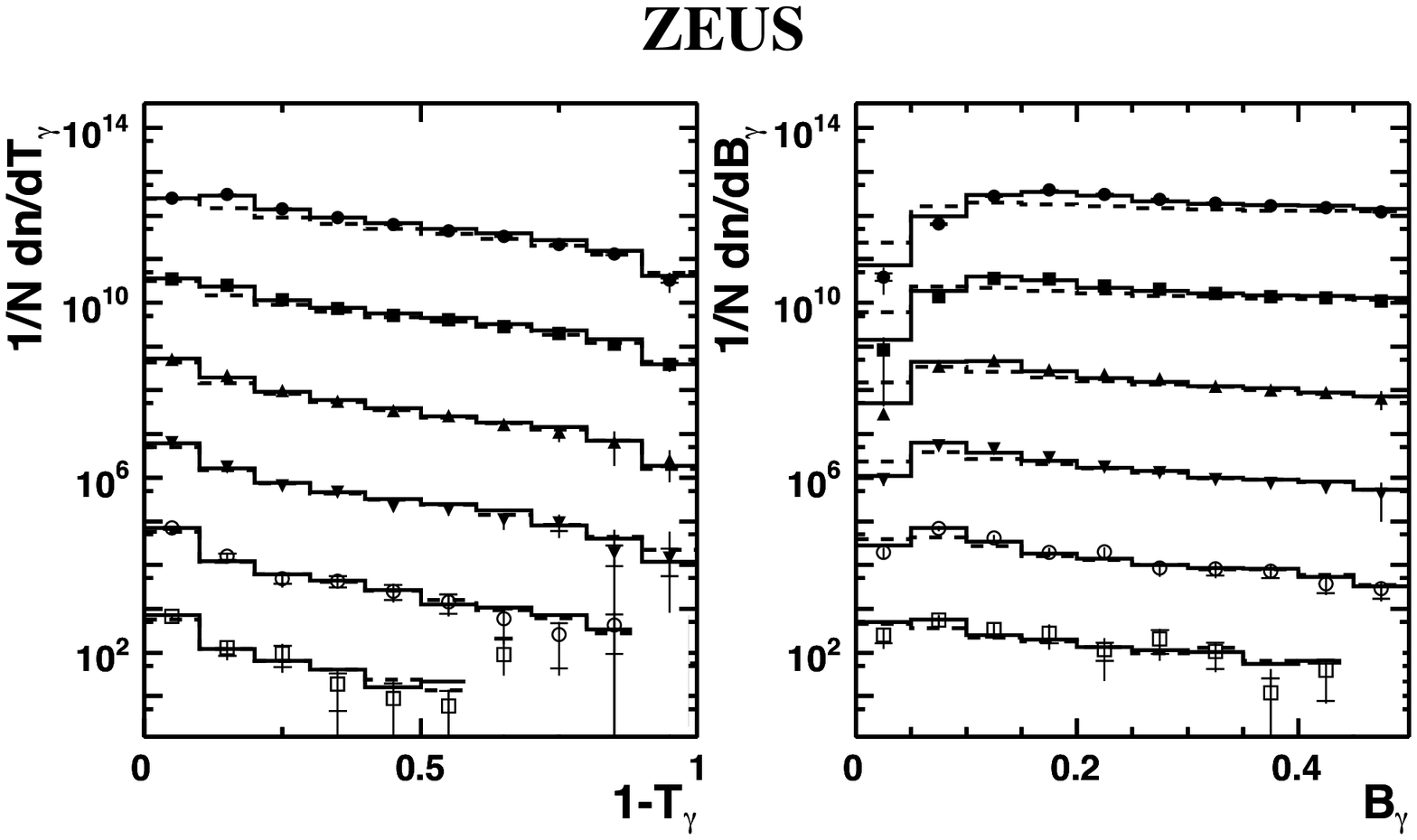,width=\linewidth}}
\end{center}
\caption{Differential distributions for the event shapes    
$1-T_\gamma$ and $B_\gamma$. Other details as in Fig.~\protect\ref{fig2new}.
}
\label{fig2newb}
\vfill
\end{figure}


\begin{figure}[htbp]
\vfill
\begin{center}
{\epsfig{figure=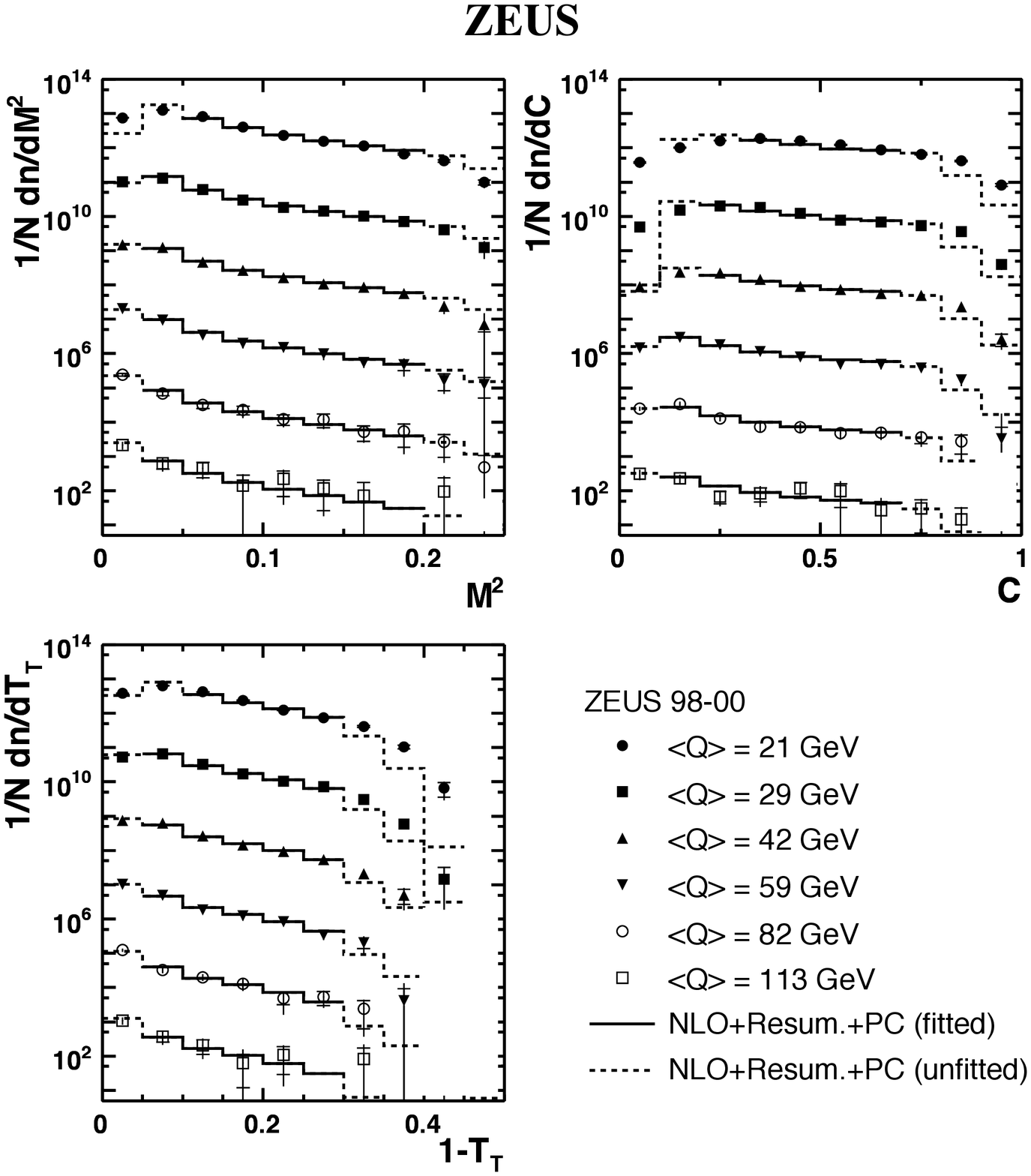,width=\linewidth}}
\end{center}
\caption{Differential distributions for the event shapes   
$M^2$, $C$ and $1-T_T$.
The distributions are normalised such that n refers to
the number of events in the $(x,Q^2)$ bin after the
$\mathcal{E}_{\rm lim}$ cut and N to the total number of
events in the $(x,Q^2)$ bin before the
$\mathcal{E}_{\rm lim}$ cut.
The differential cross section has been scaled 
for clarity by factors $10^n$, where $n$= 12, 10, 8, 6, 4, 2 for 
$<Q>$= 21, 29, 42, 59, 82 and 113 GeV, respectively. 
 The solid(dashed)  curves show the points used (omitted)
in the fit to NLL resummed calculation matched to NLO plus
power corrections.
}
\label{fig3}
\vfill
\end{figure}


\begin{figure}[htbp]
\vfill
\begin{center}
{\epsfig{figure=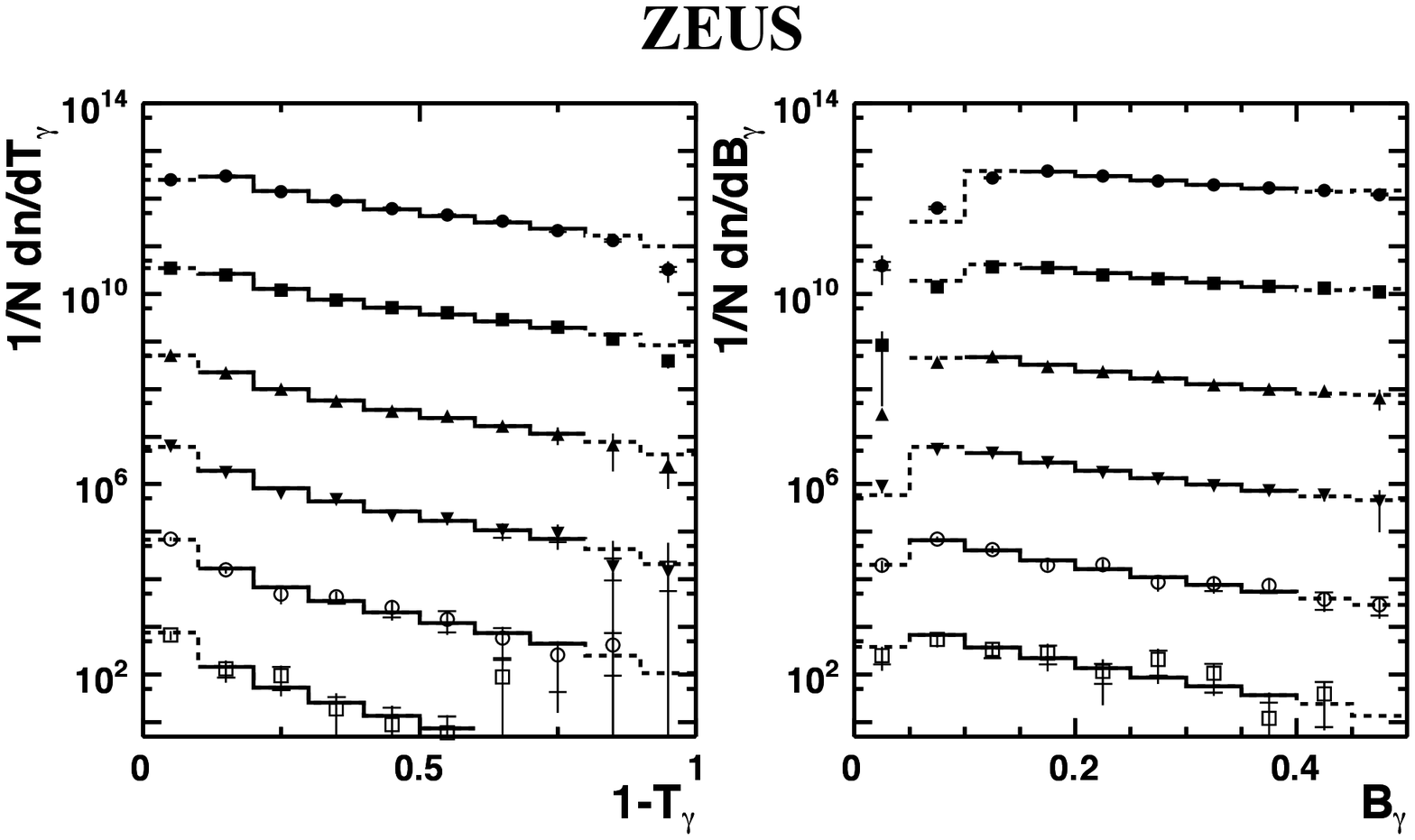,width=\linewidth}}
\end{center}
\caption{Differential distributions for the event shapes     
$1-T_\gamma$ and $B_\gamma$. Other details as in Fig.~\protect\ref{fig3}.
}
\label{fig3b}
\vfill
\end{figure}


\begin{figure}[htbp]
\vfill
\begin{center}
{\epsfig{figure=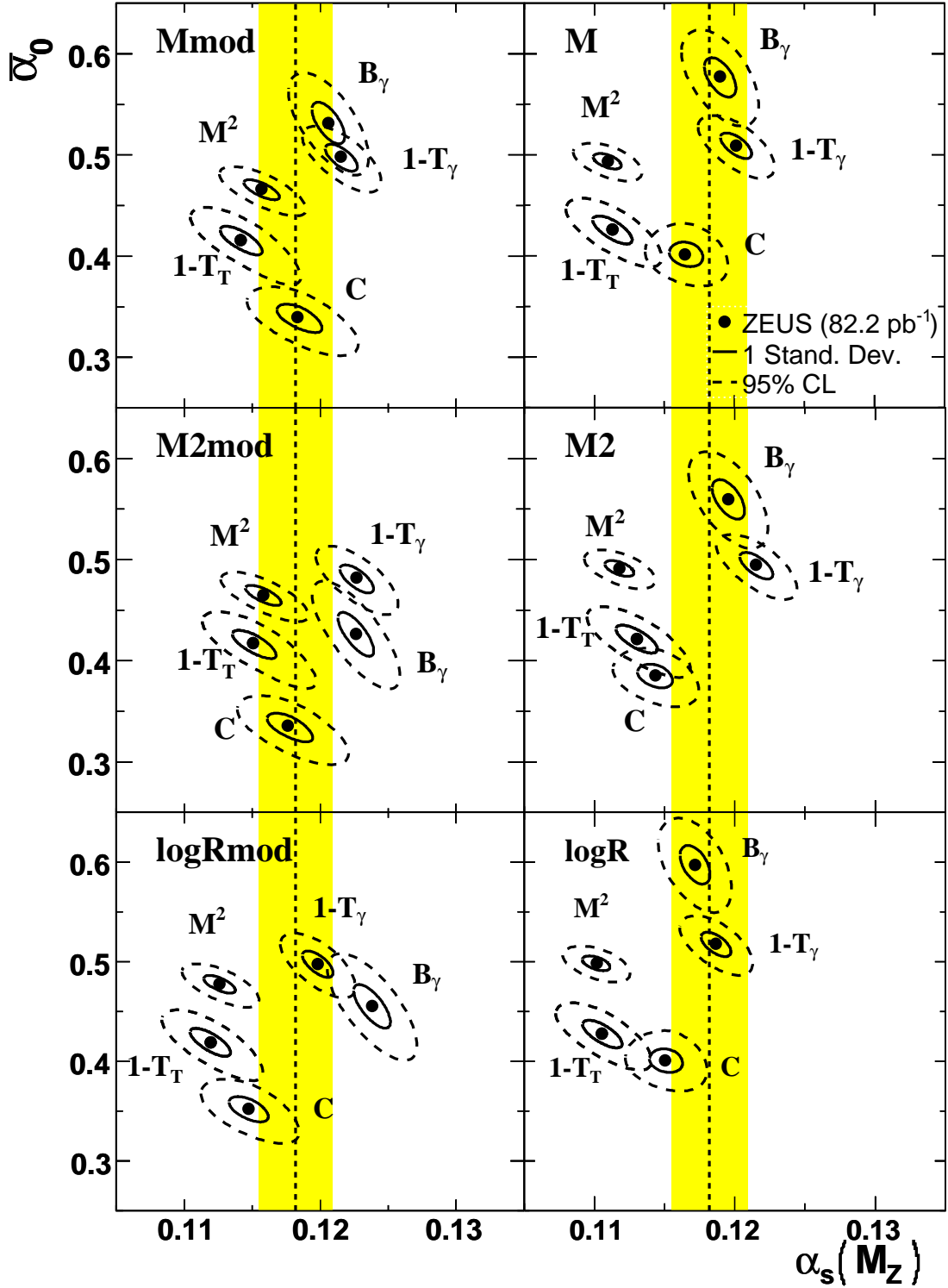,width=15cm}}
\end{center}
\caption{
        Extracted parameter values
        for $(\als,\alnot)$ from fits to differential distribution
        of shape variables  for the different matching techniques, which
	are discussed in Section~\ref{sec:pc}.
}
\label{fig4}
\vfill
\end{figure}

\begin{figure}[htbp]
\vfill
\begin{center}
\centerline{\epsfig{figure=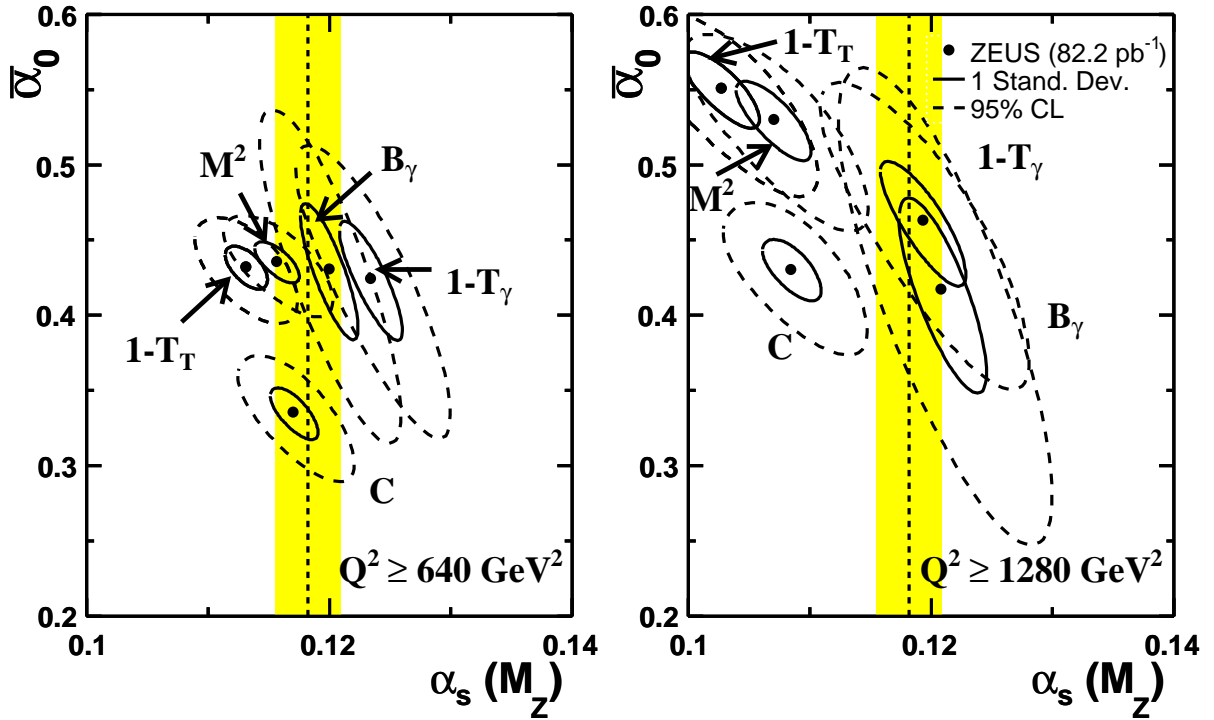,width=\linewidth}}
\end{center}
\caption{
  Extracted parameter values
  for $(\als,\alnot)$ from fits to differential distribution
  of shape variables  using  M2mod matching for
  two different $Q^2$ intervals.  }
\label{fig6}
\vfill
\end{figure}

\begin{figure}[htbp]
\vfill
\begin{center}
\psfigadd{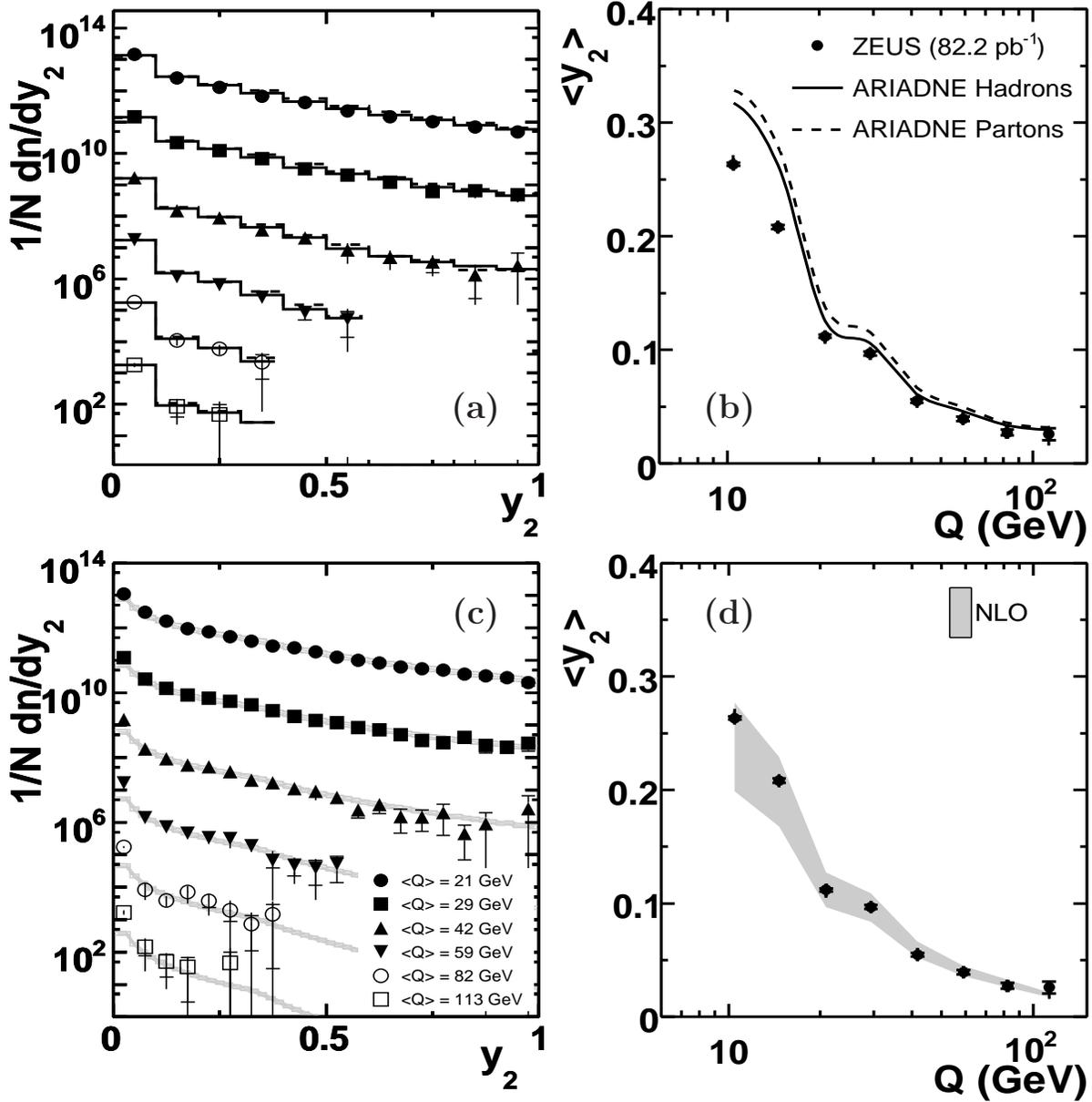}{\textwidth}{17cm}{%
                   \Text(690,1010)[]{\large\bf (a)}
		   \Text(1060,1010)[]{\large\bf (b)}
		   \Text(690,710)[]{\large\bf (c)}
                   \Text(1060,710)[]{\large\bf (d)}}
\end{center}
\caption{
(a), (c) Differential distribution for $y_2$, scaled for clarity by
different factors for each $<Q>$-value, and (b), (d) mean distribution
of $y_2$ versus $Q$ compared with predictions from (a), (b) ARIADNE
and (c), (d) NLO QCD calculations using DISENT. 
The ARIADNE predictions are shown
for the hadron (solid) and parton (dashed) levels. The NLO calculation
(shaded band) uses the CTEQ4A3 proton PDFs.
}
\label{fig7}
\vfill
\end{figure}

\begin{figure}[htbp]
\vfill
\begin{center}
\psfigadd{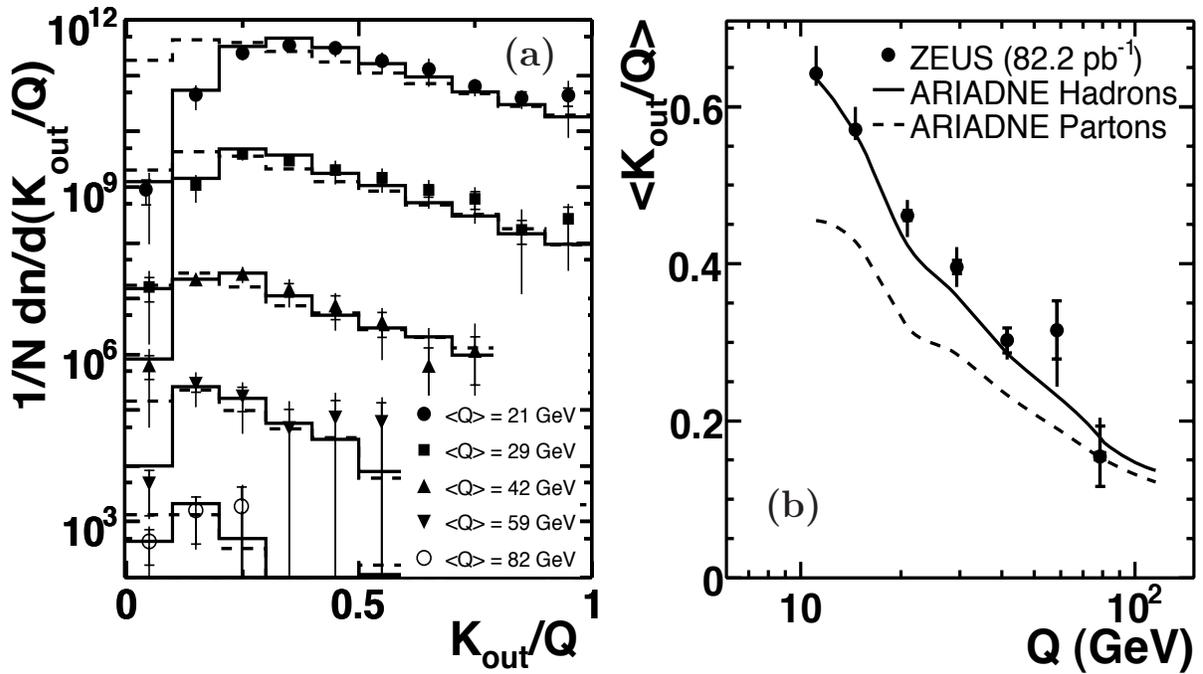}{\textwidth}{10cm}{%
                   \Text(700,830)[]{\large\bf (a)}
                   \Text(1050,230)[]{\large\bf (b)}}
\end{center}
\caption{
(a) Differential distribution for $K_{\rm OUT}/Q$, scaled for clarity
by different factors for each $<Q>$ value, and (b) mean distribution of
$K_{\rm OUT}/Q$ versus $Q$, compared with predictions from ARIADNE
for the hadron (solid) and parton (dashed) levels.
}
\label{fig11}
\vfill
\end{figure}